\documentclass[final,5p,times]{elsarticle}
\usepackage{graphicx}
\usepackage{dcolumn}
\usepackage{bm}
\usepackage[hidelinks,colorlinks=true,linkcolor=blue,citecolor=blue]{hyperref}
\usepackage{braket}
\usepackage{color}
\usepackage{float}
\usepackage{amsmath}
\usepackage[dvipsnames]{xcolor}
\usepackage{amssymb}  

\biboptions{sort&compress}

\DeclareMathOperator{\Tr}{Tr}

\begin{document}

\begin{frontmatter}

\title{Nonlinear Dissipation and Hopf Criticality in Driven Dissipative Collective Spins}

\author[addr1]{Shu Yang}
\author[addr2,addr1]{Jun Wang}
\author[addr1]{Weidong Li}
\author[addr1]{Cangtao Zhou}
\author[addr3,addr4]{Jian-Song Pan\corref{cor1}}
\ead{panjsong@scu.edu.cn}
\author[addr1]{Jianwen Jie\corref{cor2}}
\ead{Jianwen.Jie1990@gmail.com}

\cortext[cor1]{Corresponding author.}
\cortext[cor2]{Corresponding author.}

\address[addr1]{Shenzhen Key Laboratory of Ultraintense Laser and Advanced Material Technology, Center for Intense Laser Application Technology, and College of Engineering Physics, Shenzhen Technology University, Shenzhen 518118, China}
\address[addr2]{Department of Physics, Renmin University of China, Beijing 100872, China}
\address[addr3]{College of Physics, Sichuan University, Chengdu 610065, China}
\address[addr4]{Key Laboratory of High Energy Density Physics and Technology of Ministry of Education, Sichuan University, Chengdu 610065, China}

\journal{Chaos, Solitons \& Fractals}

\begin{abstract}
Self-sustained oscillations combine finite-amplitude stabilization with a neutral phase degree of freedom. We develop this bifurcation-based framework for driven-dissipative collective spins and show that the microscopic structure of the $U(1)$-covariant dissipation selects the background attractor, while the explicit $U(1)$-breaking channel governs its local bifurcation response. In the thermodynamic-limit mean-field dynamics, a single linear $U(1)$-covariant jump produces only polar fixed-point backgrounds, whereas nonlinear covariant dissipation provides amplitude-dependent saturation and stabilizes a finite-latitude self-sustained-oscillator manifold through a supercritical Hopf bifurcation. Under coherent $U(1)$ breaking, exact resonance leads to a reversible double-zero degeneracy with vanishing critical frequency rather than a standard Hopf onset. Finite detuning unfolds this singularity into a genuine finite-frequency Hopf boundary, which exists only on the self-sustained-oscillator side and can be either
supercritical or subcritical. By contrast, a single linear dissipative
$U(1)$-breaking jump cannot generate a standard Hopf instability: when
its phase-pinning invariant vanishes the azimuthal direction remains
neutral, whereas otherwise the phase-locked fixed points have a purely
real Jacobian spectrum. These results establish a general design
principle: nonlinear covariant dissipation selects the self-sustained
background, while the structure of the symmetry-breaking channel
determines whether the resulting local response is double-zero,
genuinely Hopf, or non-Hopf.
\end{abstract}


\end{frontmatter}

\section{Introduction}
Self-sustained oscillations are a canonical manifestation of
nonequilibrium nonlinear dynamics. Their emergence relies on a balance
between linear instability and nonlinear saturation: small-amplitude
oscillations are amplified, while nonlinear damping suppresses their
growth at larger amplitudes and stabilizes a finite-amplitude
oscillatory state. This amplitude-selection mechanism underlies the
van der Pol and Stuart--Landau paradigms~\cite{strogatz2024nonlinear,
pikovsky2001synchronization}. Nonlinear dissipative dynamics provides a
common framework for describing limit cycles and Hopf bifurcations in
phenomena ranging from quantum synchronization and collective
oscillations to time-crystalline order~\cite{Arinushkin2021NonlinearDamping}.
In open quantum systems,
dissipative jump processes offer a microscopic route that can control
gain and nonlinear damping and, when symmetry breaking is introduced,
the phase response as well~\cite{galve2017quantum,PRA2013Spin,
PRL2020exp,QVDP_2025_Lin}. Quantum synchronization, in particular, has
been extensively studied in
oscillators~\cite{PRL2013VdP,PRL2014VdP,PRL2019VdP,PRE2024Sudler,PRL2018VdP},
cavity and optomechanical systems~\cite{PRL2013OM1,PRA2015QED,Weiss_2016,PRE2012OM,PRA2014OM,PRL2012NOexp,PRB2009QED},
atomic ensembles~\cite{PRA2018coldatom,PRR2020hybrid}, and spin
systems~\cite{PRL2018Spin1,PRA2020two,PRL2018QN,Zhang2023PRR,PRA2022nuclear,PRR2020SPin1,Solanki2023PRA,PRA2024Tobia,PRB2024No_Go,Tan2022halfintegervs}, with
connections to entanglement~\cite{he_entanglement_2024,46my-41ym},
metrology~\cite{PRA2025QSandQFI,shen_fisher_2023},
topology~\cite{wachtler2023topological,wachtler2024topological,mr1f-v8cv},
and nonreciprocal dynamics~\cite{NC2025lai}.

Driven-dissipative collective spins provide a particularly economical
setting for analytically separating background-attractor selection from
symmetry-breaking phase control and determining how the two jointly
shape the local bifurcation structure. In the thermodynamic limit, a
$U(1)$-covariant Liouvillian may be restricted to polar fixed-point (PFP)
attractors or, with suitable nonlinear dissipation, additionally support
a self-sustained oscillator (SSO), which forms a
finite-latitude periodic orbit in the laboratory frame and a
phase-neutral stationary manifold in the co-rotating
frame~\cite{PRR2020SPin1,Solanki2023PRA,Wang2026}. Explicit
$U(1)$ breaking introduces a preferred phase reference. When phase
locking occurs, the phase-neutral manifold is reduced to isolated
rotating-frame stationary states, whose stability may change through
distinct local bifurcations. Their destabilization
can generate persistent oscillatory states and limit
cycles~\cite{Liu2026SolitonSynchronization,Duan2026PumpControlled}, thereby
connecting phase locking, collective oscillations, and time-crystalline
order~\cite{PRL2012CTC,PRL2012QTC,Montenegro2023,RMP2023Norman} within
a common bifurcation framework. The corresponding quantum critical
dynamics is reflected in the low-lying Liouvillian spectrum
through slow oscillatory branches and critical spectral
collapse~\cite{PRL2025Model,PRB2023Wang}.

This perspective is particularly relevant to persistent collective oscillations
in boundary time crystals~\cite{PRL2018BTC,PRB2021dbtc,Yang2025}. In our
previous work, we formulated a background-attractor criterion for
non-resonant boundary time crystals: detuning-robust oscillations require
an SSO-supporting background, whereas a background restricted to PFP
attractors is insufficient~\cite{Wang2026}. That criterion identifies the
required background structure, but leaves open its microscopic
dissipative origin and the bifurcation mechanisms by which coherent or
dissipative symmetry breaking reorganizes it. These open issues motivate
three questions: What is the simplest $U(1)$-covariant dissipative
structure capable of stabilizing a finite-amplitude SSO background?
How does the local bifurcation structure
under coherent symmetry breaking differ between exact resonance and
finite detuning? Can a single linear dissipative symmetry-breaking
channel generate a finite-frequency Hopf bifurcation?

Here we answer these questions by combining symmetry constraints,
mean-field flow on the Bloch sphere, linear stability analysis, and
local bifurcation analysis. First, we show that a single linear
$U(1)$-covariant jump belongs to a definite symmetry-charge sector and
can generate only polar relaxation or dephasing. It therefore cannot
select a finite-amplitude SSO background. A simple nonlinear covariant
channel, in contrast, provides the amplitude-dependent saturation
required to stabilize the SSO and generates a supercritical Hopf
transition from the PFP. This identifies a microscopic dissipative
mechanism for selecting the oscillatory background. Second, we
determine how the background responds to coherent $U(1)$ breaking. At
exact resonance, the relevant stationary branch emerges through a
double-zero degeneracy with vanishing critical frequency rather than a
standard Hopf bifurcation. At finite detuning, a genuine Hopf boundary
with nonzero critical frequency appears, but only on the SSO side of the
PFP-to-SSO transition. Third, we examine dissipative $U(1)$ breaking by
a single linear jump. Its azimuthal dependence is controlled by a single
first-harmonic quantity $Q$. When $Q=0$, the phase remains neutral. For
$Q\neq0$, any phase-locked stationary solution satisfies
$\phi^*+\psi_1=n\pi$, where the linearized latitude and phase
fluctuations decouple and the Jacobian spectrum is purely real. Thus a
single linear dissipative symmetry-breaking channel cannot produce a
finite-frequency Hopf bifurcation.

The resulting classification shows how background-selecting
dissipation and the structure of the symmetry-breaking channel jointly
determine the local bifurcation scenario. It also extends our previous
oscillatory-background criterion for non-resonant boundary time
crystals~\cite{Wang2026} by identifying both the dissipative mechanism
that selects the required SSO background and the local bifurcations
induced by symmetry breaking. The paper is organized as follows.
Section II analyzes the dissipative selection of the PFP and SSO
backgrounds, Section III examines coherent symmetry breaking at
resonance and finite detuning, and Section IV proves the absence of a
finite-frequency Hopf bifurcation for a single linear dissipative
symmetry-breaking channel. Section V summarizes the dynamical
classification and its implications. Technical derivations and
additional stability analyses are provided in the Appendices.

\section{Background attractors: PFP and SSO}
\label{sec:dissipative_structure}

The purpose of this section is structural.
We identify the minimal symmetry and dynamical requirements
for the emergence of an SSO background
in collective spin systems.
Our central result is that nonlinear,
$U(1)$-covariant dissipation is structurally necessary
for SSO backgrounds.
The discussion is organized in three steps:
we first formulate the background problem at the symmetry and
mean-field levels,
then show that linear $U(1)$-covariant dissipation yields only PFP
backgrounds,
and finally identify the nonlinear $U(1)$-covariant mechanism that
produces an SSO background through a supercritical Hopf bifurcation.

\subsection{Structural setup and criteria}
\label{subsec:background_setup}

We consider a Markovian collective-spin system governed by a Lindblad master equation
\begin{equation}
\frac{d{\rho}}{dt}
=
\mathcal{L}[{\rho}]
=
\mathcal{L}_0[{\rho}]
+
\mathcal{L}_1[{\rho}],
\label{eq:lindblad_decomposition}
\end{equation}
where the decomposition is defined according to symmetry. The system consists of $N$ identical spin-$1/2$ particles
forming a collective spin $S=N/2$,
\begin{equation}
S^\alpha=\sum_{i=1}^{N}\sigma_i^{(\alpha)}/2,
\qquad
\alpha=x,y,z .
\label{eq:collective_spin}
\end{equation}
The background Hamiltonian is taken as
\begin{equation}
H_0=\omega_0 S^z ,
\label{eq:H0_definition}
\end{equation}
which is invariant under $U(1)$ rotations generated by $S^z$,
\begin{equation}
U(\chi)=e^{-i\chi S^z}.
\label{eq:U1_rotation}
\end{equation}
We define $\mathcal{L}_0$ as the $U(1)$-covariant background
Liouvillian. It consists of $H_0$ together with dissipative processes
whose jump operators preserve the $U(1)$ symmetry generated by $S^z$.
Physically, $\mathcal{L}_0$ therefore does not introduce any external
phase reference. The remaining contribution $\mathcal{L}_1$ contains all explicit
$U(1)$-breaking terms, including both coherent and dissipative
terms. Such perturbations introduce a preferred azimuthal direction
and are treated as symmetry-breaking perturbations
acting on top of the background dynamics. This symmetry-based
decomposition is essential to the background-attractor picture adopted
here. Requiring $\mathcal{L}_0$ to be $U(1)$-covariant ensures that the
background dynamics contains no preferred azimuthal direction and may
therefore support a phase-neutral attractor manifold, while all
phase-selecting terms are assigned to $\mathcal{L}_1$. Otherwise, the
background dynamics would itself pin the azimuthal phase, making it
impossible to distinguish a phase-neutral background from its subsequent
phase locking.

A PFP background corresponds to an isolated
pole fixed point with vanishing transverse amplitude. By contrast, an
SSO background corresponds to a finite-latitude, phase-neutral
attractor manifold generated by continuous precession. The next question is therefore structural:
within the class of $U(1)$-covariant backgrounds, what kind of
dissipative mechanism yields only pole attractors, and what kind can
support a finite-latitude phase-neutral manifold?

In the thermodynamic-limit dynamics, we introduce the normalized collective magnetizations $m_\alpha=\langle S^\alpha\rangle/S,\alpha=x,y,z$, and impose the Bloch-sphere constraint $|\mathbf m|\le 1$ in the classical limit $N\to\infty$. Then, it is convenient to parametrize the Bloch sphere
by spherical coordinates
\begin{equation}
m_x=\sin\theta\cos\phi,~m_y=\sin\theta\sin\phi, ~
m_z=\cos\theta .
\label{eq:bloch_spherical}
\end{equation}
In angular variables, an SSO background requires two ingredients.
First, the latitude $\theta$ must admit a stable finite solution,
\begin{equation}
0<\theta^\ast<\pi ,
\label{eq:ssol_latitude_condition}
\end{equation}
in the absence of symmetry-breaking perturbations, so that the background
possesses finite transverse amplitude.
Second, the azimuthal angle $\phi$ must remain a neutral degree of freedom,
reflecting the $U(1)$ covariance of the background Liouvillian $\mathcal{L}_0$;
equivalently, in the rotating frame an SSO background forms a
one-parameter family of stationary states at fixed latitude. We now show that linear $U(1)$-covariant dissipation
cannot generate such a structure.
%------------------------------------------------------------
\subsection{Linear $U(1)$-covariant dissipation gives only PFP}
\label{subsec:linear_no_sso}
%------------------------------------------------------------
Consider a general linear jump operator
\begin{equation}
L
=
\sqrt{\frac{\Gamma}{S}}
\left(
a S_+
+
b S_-
+
c S_z
\right),
\label{eq:linear_jump_general}
\end{equation}
with complex coefficients $a,b,c$. Under the $U(1)$ rotation defined in
Eq.~(\ref{eq:U1_rotation}),
the spin operators transform as
\begin{equation}
S_\pm \rightarrow e^{\pm i\chi}S_\pm,
\qquad
S_z \rightarrow S_z .
\label{eq:U1_spin_transformation}
\end{equation}
\ref{app:U1_covariant_linear_jump} derives the corresponding
covariance condition explicitly.
Equations~(\ref{eq:app_component_conditions})--(\ref{eq:app_eta_c})
show that a single linear jump operator generates a
$U(1)$-covariant dissipator only if it carries a definite
$U(1)$ charge. For the general form in
Eq.~(\ref{eq:linear_jump_general}), this implies that at most one among
$a$, $b$, and $c$ can be nonzero, so that the only $U(1)$-covariant linear
channels are those proportional to $S_+$, $S_-$, or $S_z$.

Within this description, these definite-charge channels generate only polar
relaxation and do not provide the amplitude-dependent self-saturation
required to stabilize a finite-latitude orbit. As shown in
\ref{app:U1_covariant_linear_jump}, Eq.~(\ref{eq:app_theta_linear_dynamics}), the latitude dynamics takes
one of the three forms
\begin{equation}
\dot{\theta}
=
\begin{cases}
-\Gamma |a|^2\sin\theta,
&L\propto S_+,\\[4pt]
\Gamma |b|^2\sin\theta,
&L\propto S_-,\\[4pt]
0,
&L\propto S_z,
\end{cases}
\label{eq:theta_linear_dynamics}
\end{equation}
This flow admits no
interior attracting zero and therefore no stable finite-latitude
solution. Hence the generic background attractors are PFP states.

Linear $U(1)$-covariant dissipation therefore cannot generate an SSO
background: it preserves azimuthal neutrality but does not provide the
amplitude-dependent self-saturation needed for a stable finite-latitude
attractor. Guided by
this structural conclusion, in the next section we consider the
minimal nonlinear $U(1)$-covariant model introduced in
Ref.~\cite{PRL2025Model}, which provides the
required amplitude stabilization without introducing an
external phase reference, and then analyze how explicit
symmetry-breaking perturbations acting on top of this
background drive Hopf instabilities and dynamical phase
transitions.

%------------------------------------------------------------
\subsection{Nonlinear $U(1)$-covariant dissipation produces SSO}
\label{sec:III_A}
%----------------------------------------------------------
Here we consider the minimal model, whose background Liouvillian is
\begin{equation}
\mathcal{L}_0[\rho]
=
-i[H_0,\rho]
+
\mathcal D[L_+]\rho
+
\mathcal D[L_-]\rho,
\end{equation}
where $L_+$ is linear gain and $L_-$ is nonlinear damping~\cite{PRL2025Model}, given by 
\begin{equation}
L_+=\sqrt{\frac{\Gamma_+}{S}}\,S_+,
\qquad
L_-=\sqrt{\frac{\Gamma_-}{S^3}}\,S_- S_z.
\label{eq:jump_background}
\end{equation}
Under the $U(1)$ rotation defined in Eq.~(\ref{eq:U1_rotation}),
both jump operators $L_{\pm}$ transform covariantly up to phase factors.
Therefore $\mathcal{L}_0$ is $U(1)$-covariant and does not
introduce any external phase reference.

Using the spherical parametrization of Eq.~(\ref{eq:bloch_spherical}) in the laboratory frame, the background Bloch equations become
\begin{equation}
\dot\theta=-\sin\theta\bigl(\Gamma_+ - \Gamma_- \cos^{2}\theta\bigr),
\qquad
\dot\phi=\omega_0.
\label{eq:background_flow}
\end{equation}
The nonlinear channel \(L_-\) therefore provides amplitude-dependent radial stabilization, while the phase precesses uniformly at the background frequency \(\omega_0\). Besides the PFP solutions at \(\theta=0,\pi\), a finite-latitude SSO solution exists when
\begin{equation}
\cos^2\theta^\ast=\frac{\Gamma_+}{\Gamma_-},
\label{eq:background_latitude}
\end{equation}
which requires \(0<\Gamma_+/\Gamma_-<1\). Hence, for
\(\Gamma_->\Gamma_+\), two finite-latitude oscillatory branches appear.
The northern branch is attracting, whereas the southern branch is repelling.
Importantly, the emergence of the stable SSO does not eliminate all polar
attractors: in the same parameter regime the north-pole PFP is unstable,
while the south-pole PFP remains stable. The SSO therefore coexists with
a stable polar attractor. The stable northern finite-latitude branch
constitutes the SSO background.
To organize the parameter dependence, we introduce the dissipation
imbalance
\begin{equation}
\eta=\frac{\Gamma_- - \Gamma_+}{\Gamma_- + \Gamma_+},
\qquad
\Gamma_\Sigma=\Gamma_++\Gamma_-,
\label{eq:eta_parametrization}
\end{equation}
so that \(\Gamma_\pm=\Gamma_\Sigma(1\mp\eta)/2\). For brevity, we refer
to \(\eta<0\) as the PFP-only side and to \(\eta>0\) as the
SSO-supporting side. The latter terminology indicates the existence of
a stable SSO and does not exclude coexistence with the stable south-pole
PFP~\cite{PRL2025Model,Wang2026,Jie2026SingleSpin}.

%----------------------------------------------------------
\subsection{Trace-determinant classification near a stationary point}
\label{subsec:trace_determinant_classification}
%----------------------------------------------------------
Before analyzing the PFP-to-SSO transition, it is useful to classify
the linearized two-dimensional flow near a stationary point. Consider a
reduced phase-space dynamics \(\dot X = F(X)\), with
\(X=(\theta,\phi)^T\), and let \(X^\ast=(\theta^\ast,\phi^\ast)^T\) be
a stationary point,
\(F(X^\ast)=0\). Linearizing around \(X^\ast\) gives
\begin{equation}
\delta\dot X = J\,\delta X,
\qquad
J=
\left.
\frac{\partial(F_\theta,F_\phi)}{\partial(\theta,\phi)}
\right|_{X^\ast}.
\label{eq:taxonomy_linearization}
\end{equation}
At a trace-zero threshold,
\begin{equation}
\Tr J = 0,
\label{eq:taxonomy_trace_zero}
\end{equation}
the sign of \(\det J\) organizes three distinct cases:
\begin{equation}
\left\{
\begin{aligned}
\det J < 0
\;&\Longrightarrow\;
\lambda_\pm = \pm \sqrt{-\det J},
\\
\det J = 0
\;&\Longrightarrow\;
\lambda_+ = \lambda_- = 0,
\\
\det J > 0
\;&\Longrightarrow\;
\lambda_\pm = \pm i\sqrt{\det J}.
\end{aligned}
\right.
\label{eq:taxonomy_three_cases}
\end{equation}
The first case, \(\det J<0\), gives a real eigenvalue pair of opposite
sign and therefore a saddle-type instability rather than an oscillatory
onset. The second case, \(\det J=0\), is the double-zero degeneracy
encountered in Eqs.~(\ref{eq:res_equatorial_existence_main}) and
(\ref{eq:res_equatorial_eigs_main}) at exact resonance. The third case,
\(\det J>0\), gives a purely imaginary pair at trace zero and is
therefore the linear signature of a standard Hopf threshold.

In a neighborhood of this threshold, the complex-conjugate eigenvalues
can be written as
\begin{equation}
\lambda_\pm(\mu)=\beta(\mu)\pm i\omega(\mu).
\label{eq:taxonomy_hopf_eigenvalues}
\end{equation}
At the critical point \(\mu=\mu_{\mathrm c}\), one has
\(\beta(\mu_{\mathrm c})=0\) and
\(\omega(\mu_{\mathrm c})=\sqrt{\det J(\mu_{\mathrm c})}
\equiv\omega_{\mathrm H}>0\). Thus \(\beta\) in the amplitude equation
below is precisely the real part of the critical eigenvalue pair. We
orient the control parameter so that increasing \(\mu\) drives the fixed
point through its loss of stability, namely
\(\beta'(\mu_{\mathrm c})>0\). Accordingly,
\begin{equation}
\begin{aligned}
\beta(\mu)&<0, && \mu<\mu_{\mathrm c},\\
\beta(\mu_{\mathrm c})&=0, && \mu=\mu_{\mathrm c},\\
\beta(\mu)&>0, && \mu>\mu_{\mathrm c}.
\end{aligned}
\label{eq:taxonomy_hopf_beta_sign}
\end{equation}
Small perturbations of the fixed point therefore decay before the
threshold, are marginal at the threshold, and grow after the threshold.

Once the third case holds, one can further determine whether the Hopf
threshold is supercritical or subcritical. After center-manifold
reduction and normal-form transformation, the reduced amplitude
dynamics takes the form~\cite{Kuznetsov2004}
\begin{equation}
\dot r = \beta(\mu)r + l\,r^3 + O(r^5),
\label{eq:taxonomy_hopf_normal_form}
\end{equation}
Here \(r\ge0\) is the scalar amplitude of the critical oscillatory
mode on the center manifold, \(\mu\) denotes the control parameter, and
\(\mu_{\mathrm c}\) is its critical value at the onset. The coefficient
\(\beta(\mu)\) is the linear growth rate identified above. The
coefficient \(l\) is the cubic nonlinear coefficient,
equivalently the first Lyapunov coefficient in this scalar amplitude
equation. The remainder \(O(r^5)\) collects terms of fifth and higher
order in the amplitude \(r\). If \(l<0\), the
Hopf bifurcation is supercritical and creates a stable small-amplitude
limit cycle when the fixed point loses stability. If \(l>0\), it is
subcritical and the nearby small limit cycle is unstable. Thus
supercritical versus subcritical is a nonlinear refinement of the
\(\det J>0\) case, not a separate linear class.

\begin{figure}[t]
\centering
\includegraphics[width=0.75\columnwidth]{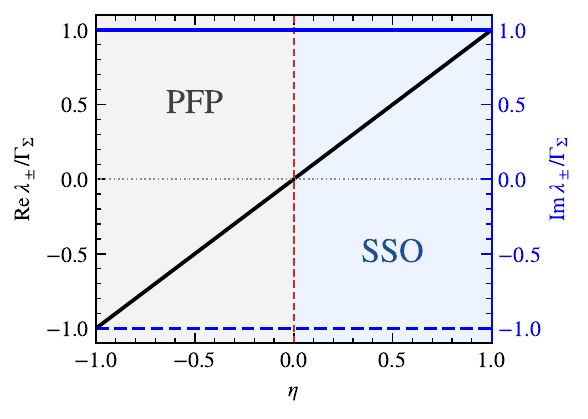}
\caption{
Eigenvalues in Eq.~(\ref{eq:background_lambda}) versus \(\eta\), shown for
\(\omega_0/\Gamma_\Sigma=1\). The black curve and left axis give
\(\operatorname{Re}\lambda_\pm/\Gamma_\Sigma\); the solid and dashed
blue curves and right axis give the two signs of
\(\operatorname{Im}\lambda_\pm/\Gamma_\Sigma\). The red dashed line
at \(\eta=0\) marks the Hopf critical point, given by its intersection
with the black curve.
}
\label{fig:background_eigenvalues}
\end{figure}

%----------------------------------------------------------
\subsection{The north-pole PFP-to-SSO transition is a supercritical Hopf bifurcation}
%----------------------------------------------------------
The model-specific Hopf criterion is most transparent in the laboratory
frame. The full Cartesian linearization is given in
\ref{app:pfp_sso_hopf}; around the north-pole PFP it yields the
transverse eigenvalues
\begin{equation}
\lambda_\pm=(\Gamma_- - \Gamma_+)\pm i\omega_0.
\label{eq:background_lambda}
\end{equation}
At \(\Gamma_-=\Gamma_+\), this pair crosses the imaginary axis at the
nonzero frequency \(\omega_0\). The north-pole PFP is therefore stable
for \(\Gamma_-<\Gamma_+\) and loses stability for
\(\Gamma_->\Gamma_+\), establishing a standard Hopf threshold.
The nonlinear type follows from the transverse amplitude
\(r=\sqrt{m_x^2+m_y^2}\). \ref{app:pfp_sso_hopf} derives
\begin{equation}
\dot r=\mu r-\Gamma_-r^3+O(\mu r^3,r^5),
\qquad
\mu=\Gamma_- - \Gamma_+.
\label{eq:background_amplitude_normal_form}
\end{equation}
The cubic coefficient \(-\Gamma_-\) is negative, so for \(\mu>0\) an
attracting nonzero branch emerges with
\(r_\ast=\sqrt{(\Gamma_--\Gamma_+)/\Gamma_-}\), as given in
Eq.~(\ref{eq:app_r_star}). Because \(r_\ast\propto\mu^{1/2}\) vanishes
continuously at onset, the bifurcation is supercritical. In the
laboratory frame this branch is a stable limit cycle, while in the
rotating frame it is the phase-neutral SSO background manifold.

%============================================================
\section{Coherent $U(1)$-breaking drive}
\label{sec:coherent_u1_breaking_drive}
%============================================================

\begin{figure*}[t]
\centering
\includegraphics[width=0.8 \textwidth]{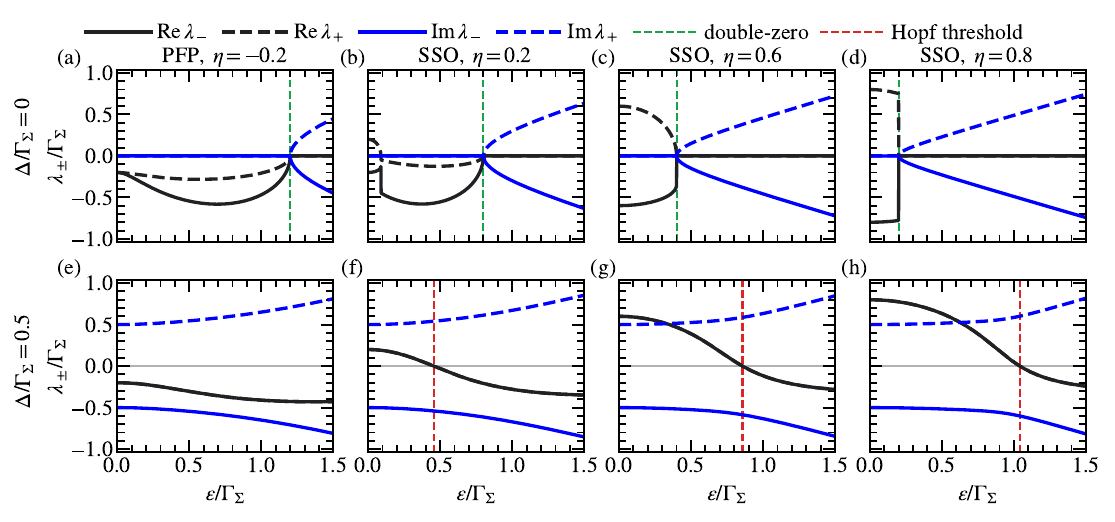}
\caption{
Eigenvalues for \(\Delta/\Gamma_\Sigma=0\) and \(0.5\) (rows) and
\(\eta=-0.2,0.2,0.6,\) and \(0.8\) (columns). Black curves
show \(\mathrm{Re}\,\lambda_{1,2}/\Gamma_\Sigma\),
blue curves show \(\mathrm{Im}\,\lambda_{1,2}/\Gamma_\Sigma\). The
green dashed lines in the resonant row mark the double-zero threshold,
while the red dashed lines in the detuned row mark the standard Hopf
threshold for \(\Delta\neq0\) from
Eq.~(\ref{eq:detuned_epsilon_bound}).
}
\label{fig:app_detuned_eigenvalue_crossing}
\end{figure*}

%----------------------------------------------------------
\subsection{Angular equations and fixed-point conditions}
%----------------------------------------------------------
We now add a coherent drive,
\begin{equation}
H_\epsilon(t)=-\epsilon\cos(\omega t)S_x,
\qquad
\mathcal{L}_1[\rho]=-i[H_\epsilon(t),\rho],
\label{eq:H_epsilon}
\end{equation}
After transforming to the frame rotating at \(\omega_0\) and applying
the rotating-wave approximation, the equations of motion become
\begin{align}
\dot\theta
&=
-\sin\theta
\Big(
\Gamma_+ - \Gamma_- \cos^{2}\theta
\Big)
+
\frac{\epsilon}{2} \sin\phi ,
\label{eq:theta_epsilon}
\\
\dot\phi
&=
\Delta
+
\frac{\epsilon}{2}
\cot\theta
\cos\phi .
\label{eq:phi_epsilon}
\end{align}
where \(\Delta=\omega-\omega_0\) is the detuning. Denoting the
right-hand sides of Eqs.~(\ref{eq:theta_epsilon}) and
(\ref{eq:phi_epsilon}) by \(F_\theta\) and \(F_\phi\), stationary fixed
points satisfy
\begin{equation}
F_\theta(\theta^\ast,\phi^\ast)=0,
\qquad
F_\phi(\theta^\ast,\phi^\ast)=0 ,
\label{eq:hopf_stationary_conditions}
\end{equation}
The fixed-point equations take qualitatively different forms at exact
resonance and at finite detuning. In particular, when \(\Delta=0\), the
phase-locking condition factorizes and gives rise to a distinct
equatorial branch, whereas for \(\Delta\neq0\) the latitude and phase
remain coupled. We therefore analyze the resonant and detuned cases
separately below.
For finite detuning, combining
Eqs.~(\ref{eq:theta_epsilon})--(\ref{eq:hopf_stationary_conditions})
gives the fixed-point relations
\begin{align}
2\Delta
&=
-\epsilon\cot\theta^\ast\cos\phi^\ast,
\label{eq:phi_detuned}\\
\epsilon^2
&=
4\sin^2\theta^\ast
\bigl(
\Gamma_+ - \Gamma_- \cos^2\theta^\ast
\bigr)^2
+
4\Delta^2\tan^2\theta^\ast .
\label{eq:detuned_fixed_point_implicit}
\end{align}
At resonance, Eq.~(\ref{eq:phi_detuned}) factorizes into two
distinct possibilities,
\begin{equation}
\cos\phi^\ast=0
\qquad\text{or}\qquad
\theta^\ast=\frac{\pi}{2}.
\label{eq:res_two_branches_main}
\end{equation}
The equatorial solution is not represented regularly by
Eq.~(\ref{eq:detuned_fixed_point_implicit}), because
\(\Delta^2\tan^2\theta^\ast\) becomes indeterminate in the simultaneous
limit \(\Delta\to0\) and \(\theta^\ast\to\pi/2\).

Linearizing around a stationary point \((\theta^\ast,\phi^\ast)\) gives
the \(2\times2\) Jacobian
\(J_\ast=\partial(F_\theta,F_\phi)/\partial(\theta,\phi)|_\ast\), whose
matrix elements are
\begin{align}
J_{11}^\ast
&=
\left.\partial_\theta F_\theta\right|_\ast
= [-\Gamma_+
+\Gamma_-\bigl(3\cos^2\theta^\ast-2\bigr)]\cos\theta^\ast,
\label{eq:main_j11}\\
J_{12}^\ast
&=
\left.\partial_\phi F_\theta\right|_\ast
=
\frac{\epsilon}{2}\cos\phi^\ast,
\label{eq:main_j12}\\
J_{21}^\ast
&=
\left.\partial_\theta F_\phi\right|_\ast
=
-\frac{\epsilon}{2}\csc^2\theta^\ast\cos\phi^\ast,
\label{eq:main_j21}\\
J_{22}^\ast
&=
\left.\partial_\phi F_\phi\right|_\ast
=
-\frac{\epsilon}{2}\cot\theta^\ast\sin\phi^\ast.
\label{eq:main_j22}
\end{align}

A Hopf
bifurcation occurs when the stationary point changes stability
through a complex-conjugate eigenvalue pair. As summarized in
Sec.~\ref{subsec:trace_determinant_classification}, the generic two-dimensional classification implies that a
standard Hopf crossing here requires
\(\Tr J_\ast=0\) together with \(\det J_\ast>0\); see
Eqs.~(\ref{eq:taxonomy_trace_zero}) and
(\ref{eq:taxonomy_three_cases}).

%----------------------------------------------------------
\subsection{Resonance: double-zero, not standard Hopf}
%----------------------------------------------------------
At resonance, the stationary equation for \(\phi\) splits the fixed
points into the two branches already indicated in
Eq.~(\ref{eq:res_two_branches_main}). The first branch,
\(\cos\phi^\ast=0\), is a real-eigenvalue branch. Its fixed-point
existence is controlled by a polynomial in
\(x=\cos^2\theta^\ast\), and its endpoint can be a saddle-node or a
branch intersection depending on the dissipation imbalance. Its
linearized spectrum is real, so its loss of existence is not a
nonzero-frequency crossing of a complex-conjugate pair.

\begin{figure*}[t]
\centering
\includegraphics[width=0.8\textwidth]{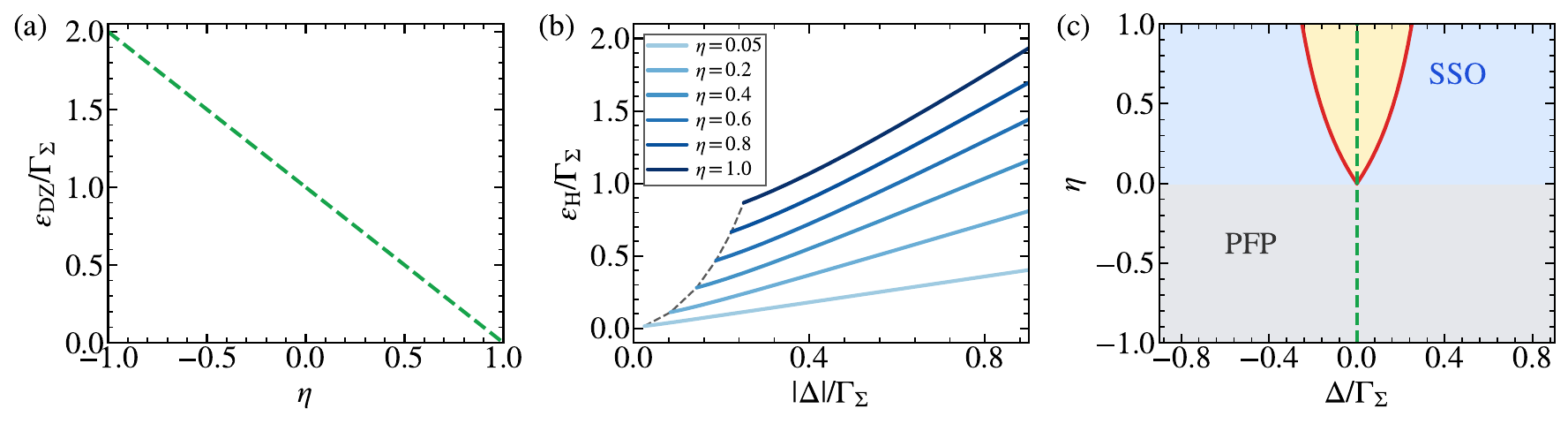}
\caption{
Threshold structure of the coherent driven system at
\(\Gamma_\Sigma=1\). Panel (a) shows the resonant
double-zero threshold \(\epsilon_{\mathrm{DZ}}/\Gamma_\Sigma\) versus
\(\eta\). Panel (b) shows standard Hopf thresholds for
\(\eta=0.05,0.2,0.4,0.6,0.8,\) and \(1.0\). The colored branches in
panel (b) are obtained from Eqs.~(\ref{eq:detuned_fixed_point_implicit}),
(\ref{eq:trace_detuned_compact}), and
(\ref{eq:det_detuned_compact}). The dashed guide traces
the left endpoints of the corresponding SSO branches. Panel (c) shows
the \((\Delta/\Gamma_\Sigma,\eta)\) plane; the green dashed line marks
the resonant double-zero locus, while the red solid boundaries mark
the finite-detuning Hopf critical lines.
}
\label{fig:detuning_phase_boundary}
\end{figure*}

The Equatorial branch is defined by
\begin{equation}
\theta^\ast=\frac{\pi}{2},
\qquad
\sin\phi^\ast=\frac{2\Gamma_+}{\epsilon},
\label{eq:res_equatorial_main}
\end{equation}
which exists for
\begin{equation}
\epsilon\ge\epsilon_{\mathrm{DZ}},
\qquad
\epsilon_{\mathrm{DZ}}\equiv2\Gamma_+ .
\label{eq:res_equatorial_existence_main}
\end{equation}
Linearizing the resonant flow on this branch gives
\begin{equation}
\lambda_\pm
=
\pm\frac{i}{2}
\sqrt{\epsilon^2-\epsilon_{\mathrm{DZ}}^2} .
\label{eq:res_equatorial_eigs_main}
\end{equation}
This resonant eigenvalue structure is shown explicitly in panels
(a)--(d) of Fig.~\ref{fig:app_detuned_eigenvalue_crossing}, namely the
first row at \(\Delta/\Gamma_\Sigma=0\) for
\(\eta=-0.2,0.2,0.6,\) and \(0.8\). In all four panels, it is the real
part that reaches zero at the dashed threshold, while the imaginary part appears only for
\(\epsilon>\epsilon_{\mathrm{DZ}}\), so the onset is a double-zero
degeneracy rather than a standard Hopf crossing with finite critical
frequency. Panel~(a), on the PFP side, shows the real-eigenvalue branch
approaching and terminating at the double-zero intersection, where the
equatorial branch emerges. Beyond the threshold, the equatorial branch
carries a purely imaginary eigenvalue pair. Panels~(b)--(d) show the
corresponding resonant double-zero onset on the SSO side. For fixed
total dissipation \(\Gamma_\Sigma\), as \(\eta\) increases,
\(\Gamma_+\) decreases and the threshold
\(\epsilon_{\mathrm{DZ}}=2\Gamma_+\) shifts leftward to lower drive
strength, consistent with the sequence from panel~(b) to panel~(d).
Thus \(\epsilon_{\mathrm{DZ}}=2\Gamma_+\) marks the onset of the
equatorial branch. At this onset the eigenvalue pair forms a
double-zero eigenvalue degeneracy in the standard trace-determinant
sense~\cite{Kuznetsov2004},
whereas for \(\epsilon>\epsilon_{\mathrm{DZ}}\) the equatorial branch carries a
purely imaginary pair.
The complete root multiplicity and branch endpoints are worked out in
\enlargethispage{2\baselineskip}
\ref{app:real_eigenvalue_branch_details}, with the corresponding
branch structure displayed in Figs.~\ref{fig:app_gx_landscape}
and~\ref{fig:app_fixed_point_branches}. The equatorial double-zero onset and
the associated nonlinear center structure are derived in
\ref{app:equatorial_branch_details}, as summarized by
Eqs.~(\ref{eq:app_resonant_dz_threshold}),
(\ref{eq:app_resonant_soft_mode}), and
(\ref{eq:app_resonant_reversibility}).

\begin{figure*}[t]
\centering
\includegraphics[width=0.85 \textwidth]{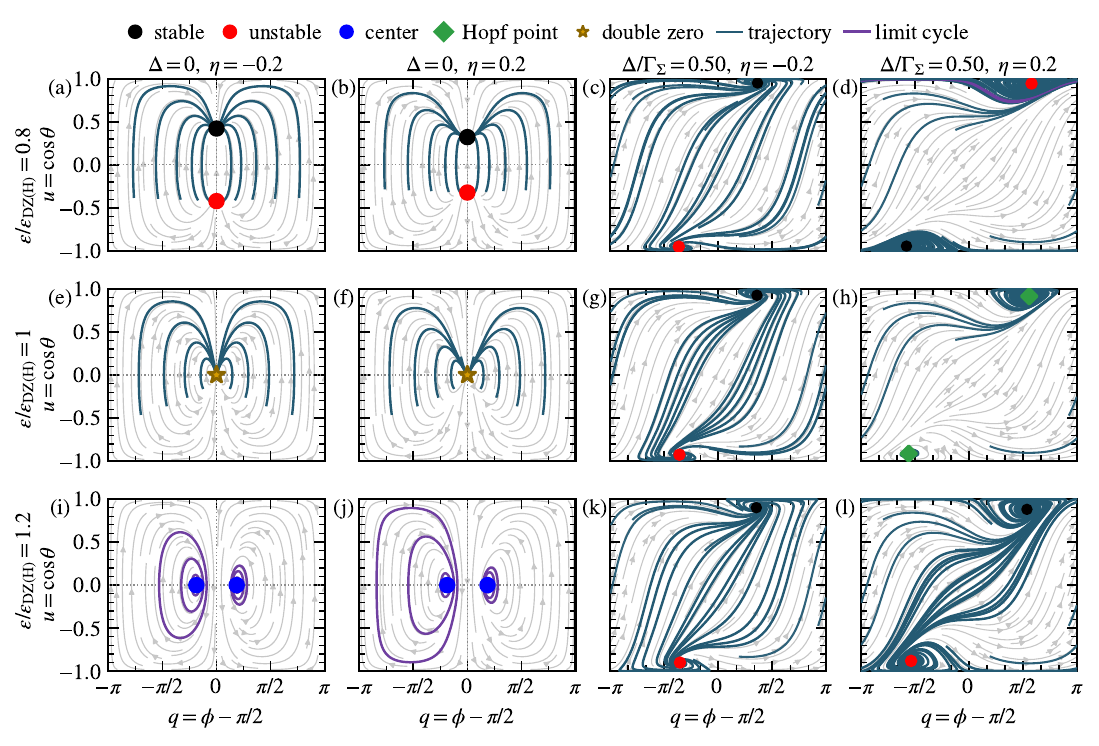}
\caption{
Phase portraits in \((q,u)=(\phi-\pi/2,\cos\theta)\) for
\(\eta=-0.2\) and \(0.2\). The left and right two columns show
\(\Delta/\Gamma_\Sigma=0\) and \(0.5\), respectively. From top to bottom,
\(\epsilon/\epsilon_{\mathrm{DZ}}=0.8,1.0,1.2\) in the resonant
columns and
\(\epsilon/\epsilon_{\mathrm H}^{(\Delta)}=0.8,1.0,1.2\) in the
finite-detuning columns. Gray streamlines show the global flow field,
dark-teal curves highlight representative trajectories from selected
initial conditions, and purple curves highlight representative closed
periodic orbits: center orbits in the resonant panels and isolated limit
cycles in the finite-detuning panels. Filled black, red, and blue circles denote stable
fixed points, unstable fixed points, and reversible centers,
respectively. Green diamonds mark finite-detuning Hopf points, and
the gold star marks the resonant reversible double-zero point.
}
\label{fig:resonant_detuned_phase_portraits}
\end{figure*}

%----------------------------------------------------------
\subsection{Finite detuning: genuine standard Hopf only in SSO regime}
\label{sec:hopf_detuned}
%----------------------------------------------------------
For finite detuning, the question is whether an existing stationary
branch satisfies the standard Hopf conditions. Using the stationary
equations,
the trace and determinant reduce to
\begin{equation}
\Tr J
=
-2\cos\theta^\ast
\bigl(
\Gamma_+ - \Gamma_- \cos 2\theta^\ast
\bigr) .
\label{eq:trace_detuned_compact}
\end{equation}
and
\begin{equation}
\begin{aligned}
\det J
={}&
\cos^2\theta^\ast
\bigl(
\Gamma_+ - \Gamma_- \cos^2\theta^\ast
\bigr)
\\
&\quad\times
\bigl(
\Gamma_+ + 2\Gamma_- - 3\Gamma_- \cos^2\theta^\ast
\bigr)
+
\frac{\Delta^2}{\cos^2\theta^\ast} .
\end{aligned}
\label{eq:det_detuned_compact}
\end{equation}
\ref{app:finite_detuning_details} derives these expressions and
shows why the resonant equatorial branch, \(\theta^\ast=\pi/2\), is excluded at
\(\Delta\neq0\); see
Eqs.~(\ref{eq:app_phi_detuned})--(\ref{eq:app_detuned_trace_branch}).
The remaining
trace-zero branch is
\begin{equation}
\cos^2\theta^\ast=\frac{\Gamma_-+\Gamma_+}{2\Gamma_-},
\label{eq:detuned_trace_zero_branch}
\end{equation}
which is physical only for \(\Gamma_->\Gamma_+\), namely in the SSO
regime. Thus a standard Hopf solution at finite detuning is absent in the PFP
regime, because the PFP background lacks the interior trace-zero
stationary branch needed for a nonzero-frequency complex-pair
crossing. Using the parametrization in
Eq.~(\ref{eq:eta_parametrization}), the condition \(\det J>0\) gives the
dimensionless detuning bound
\begin{equation}
\frac{|\Delta|}{\Gamma_\Sigma}>
\frac{\eta}{2(1+\eta)},
\qquad \eta>0.
\label{eq:detuned_delta_bound}
\end{equation}
The corresponding drive threshold is
\begin{equation}
\left[\frac{\epsilon_{\mathrm H}(\Delta)}{\Gamma_\Sigma}\right]^2
=
\frac{\eta^3}{1+\eta}
+4\eta\left(\frac{\Delta}{\Gamma_\Sigma}\right)^2.
\label{eq:detuned_epsilon_bound}
\end{equation}
These formulas apply only to the non-resonant SSO branch satisfying
Eq.~(\ref{eq:detuned_delta_bound}); they do not extend to
\(\Delta=0\), where the onset is instead a double-zero
degeneracy.
This dependence is summarized in panel (b) of
Fig.~\ref{fig:detuning_phase_boundary}, which plots
\(\epsilon_{\mathrm H}/\Gamma_\Sigma\) versus
\(|\Delta|/\Gamma_\Sigma\) for representative \(\eta>0\). Each solid
curve begins only beyond the corresponding minimum detuning set by
Eq.~(\ref{eq:detuned_delta_bound}), and then rises with increasing
\(|\Delta|\). As \(\eta\) increases, the finite-detuning Hopf threshold
shifts rightward to higher drive strength, showing that deeper SSO
backgrounds require a stronger coherent drive to reach the Hopf
instability.
This distinction is visible in panels (e)--(h) of
Fig.~\ref{fig:app_detuned_eigenvalue_crossing}, which show the detuned
spectra at \(\Delta/\Gamma_\Sigma=0.5\) for
\(\eta=-0.2,0.2,0.6,\) and \(0.8\). Panel (e), on the PFP side, never
develops a trace-zero complex-pair crossing: the real parts do not pass
through zero in the Hopf form, consistent with the absence of a
finite-detuning standard Hopf threshold in that regime. By contrast,
panels (f)--(h) lie on the SSO side and display the genuine detuned
Hopf scenario: the real part of the complex pair crosses zero at the
threshold while the imaginary part remains nonzero. The Hopf threshold
shifts to higher drive strength as \(\eta\) increases from panel~(f) to
panel~(h), in agreement with Eq.~(\ref{eq:detuned_epsilon_bound}).

The broader threshold organization is summarized by panels (b) and (c)
of Fig.~\ref{fig:detuning_phase_boundary}. Panel (b) collects the
finite-detuning Hopf thresholds as functions of \(|\Delta|\), so each
solid branch begins only beyond the minimum detuning allowed by
Eq.~(\ref{eq:detuned_delta_bound}) and then rises with increasing
\(|\Delta|\). Panel (c) places this information back into the full
\((\Delta/\Gamma_\Sigma,\eta)\) plane: the green dashed line is the
resonant double-zero locus at \(\Delta=0\), whereas the red solid
curves are the finite-detuning Hopf critical lines. Thus the PFP side
\((\eta<0)\) carries only the resonant double-zero line, while the SSO
side \((\eta>0)\) additionally supports standard Hopf thresholds away
from resonance.

\begin{figure*}[t]
\centering
\includegraphics[width=0.92\textwidth]{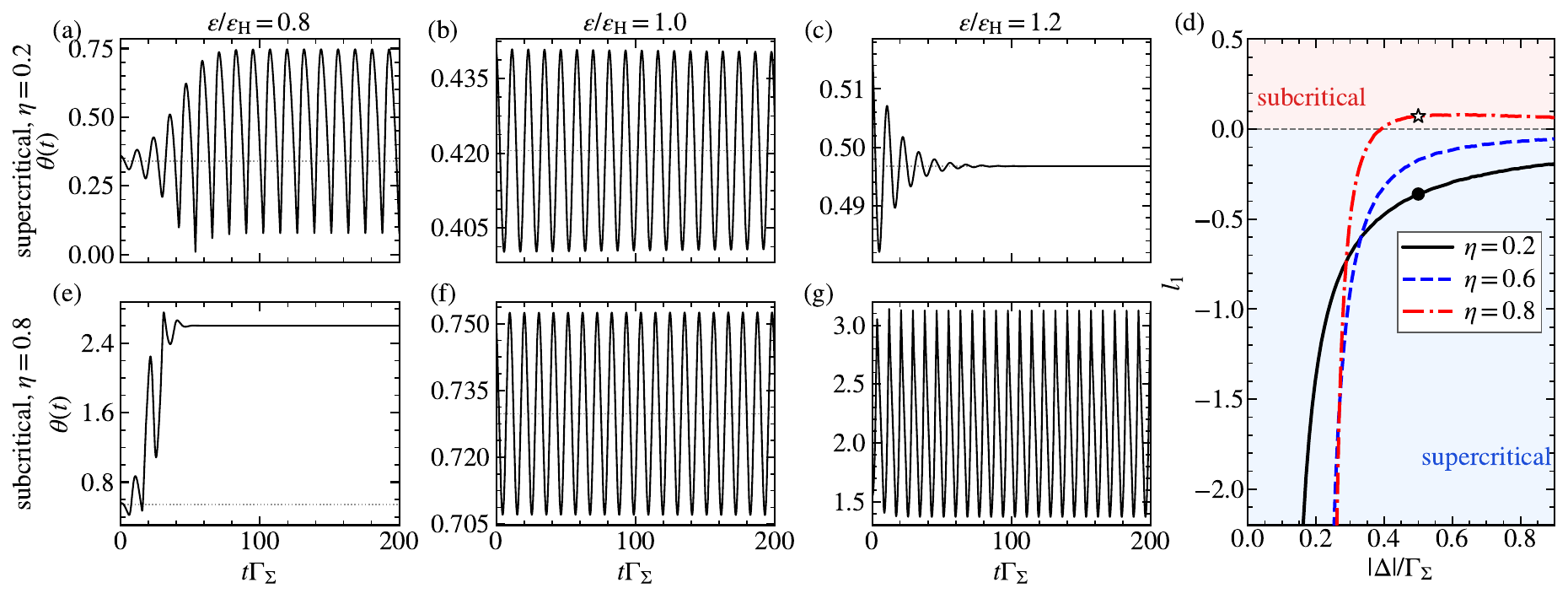}
\caption{
Detuned SSO dynamics at \(\Delta/\Gamma_\Sigma=0.5\) and nonlinear
classification along the finite-detuning Hopf branches for
representative values of \(\eta\). Panels (a)--(c) show representative
time traces \(\theta(t)\) for \(\eta=0.2\), while panels (e)--(g)
show the corresponding \(\eta=0.8\) traces. In each set,
\(\epsilon/\epsilon_{\mathrm H}=0.8,1.0,\) and \(1.2\) from left to
right, and the evolution extends to \(t\Gamma_\Sigma=200\). Panel (d)
shows the first Lyapunov coefficient \(l_1\) along representative
finite-detuning Hopf branches. The lower and upper shaded regions,
\(l_1<0\) and \(l_1>0\), denote supercritical and subcritical
bifurcations, respectively. The filled black circle and open star mark
the \(\eta=0.2\) and \(0.8\) cases shown in the time traces. The
\(\eta=0.8\) branch changes sign near
\(|\Delta|/\Gamma_\Sigma\approx0.39\).
}
\label{fig:detuned_hopf_dynamics_l1}
\end{figure*}

Figure~\ref{fig:resonant_detuned_phase_portraits} provides the phase-space
summary of Sec.~III. On resonance, the flow reorganizes at
\(\epsilon=\epsilon_{\mathrm{DZ}}\) through the reversible double-zero
onset discussed above, rather than through a standard Hopf birth of an
isolated limit cycle. The purple closed trajectories in the resonant
columns are the thermodynamic-limit phase-space manifestation of the
resonant boundary time crystal first identified in
Ref.~\cite{PRL2018BTC}. By contrast, away from resonance the PFP side
still supports only non-closed trajectories, whereas the detuned SSO
portrait reveals the non-resonant boundary time crystal:
there the closed running trajectory is sustained by the underlying SSO
background structure, precisely as emphasized in our recent
work~\cite{Wang2026}. Placing the
resonant and detuned portraits side by side therefore makes the central
distinction of this section visually explicit: the resonant onset is a
double-zero reorganization connected to the original resonant BTC,
whereas the finite-detuning BTC emerges only from the SSO side and thus
has a distinct non-resonant origin.
The nonlinear Hopf type along this finite-detuning SSO boundary is also
not uniform. This is shown in
Fig.~\ref{fig:detuned_hopf_dynamics_l1}, which combines representative
detuned SSO time traces at \(\Delta/\Gamma_\Sigma=0.5\) with the
numerical first Lyapunov coefficient \(l_1\) evaluated on the true
detuned Hopf locus. For the \(\eta=0.2\) branch, panels (a)--(c) show
that the oscillation turns on continuously as the threshold is crossed:
Above threshold the phase-locked fixed point is unstable and the dynamics approached a stable limit cycle, at threshold the dynamics exhibits critically slow relaxation, while above threshold the fixed point is stable and the trajectory relaxes toward it. Equivalently, the stable oscillation turns on continuously when \(\epsilon\) is decreased through \(\epsilon_{\mathrm H}\), which is the characteristic supercritical scenario. By contrast, panels (e)--(g) show that for \(\eta=0.8\) the
near-threshold dynamics is not captured by a small stable cycle born
continuously from the fixed point; instead, the sustained oscillation is
reached only through the finite-amplitude basin of an outer stable
cycle, which is the dynamical signature of a subcritical Hopf
bifurcation.

Panel (d) organizes these representative traces into a global nonlinear
classification. The \(\eta=0.2\) and \(0.6\) branches remain in the
\(l_1<0\) sector over the plotted detuning range and are therefore
supercritical throughout, whereas the \(\eta=0.8\) branch crosses
\(l_1=0\) near \(|\Delta|/\Gamma_\Sigma\approx0.39\) and becomes
subcritical at larger detuning. Thus the finite-detuning SSO threshold
is a genuine standard Hopf instability at the linear level, but its
nonlinear criticality depends on where the branch lies in the
\((\eta,\Delta)\) plane. The derivation of \(l_1\) is given in
\ref{app:finite_detuning_details}.

%============================================================
\section{Dissipative $U(1)$-breaking perturbation}
\label{sec:dissipative_u1_breaking}
%============================================================

%------------------------------------------------------------
\subsection{General linear dissipative perturbation}
\label{subsec:linear_dissipative_perturbation}
%------------------------------------------------------------

We next ask how the same $U(1)$-covariant background responds
to an explicitly dissipative $U(1)$-breaking perturbation.
We return to the general linear collective jump operator
introduced in Eq.~(\ref{eq:linear_jump_general}),
\begin{equation*}
L=
\sqrt{\frac{\Gamma}{S}}
\left(
aS_+ + bS_- + cS_z
\right),
\end{equation*}
with complex coefficients $a$, $b$, and $c$.
As discussed in Sec.~II B and proved in Appendix~A, a single
linear jump is $U(1)$ covariant only when it carries a definite
$U(1)$ charge. Hence any genuine mixing between different
charge sectors explicitly breaks the $U(1)$ symmetry and can
provide an azimuthal phase reference.

The thermodynamic-limit mean-field equations generated by this
general linear jump are derived in Appendix~D. Adding the jump
to the rotating-frame background dynamics gives
\begin{align}
\dot{\theta}
={}&
-\sin\theta
\left(
\Gamma_+-\Gamma_-\cos^2\theta
\right)
\nonumber\\
&-\Gamma
\left[
\alpha\sin\theta
+
|Q|\cos\theta\cos(\phi+\psi_1)
\right],
\label{eq:diss_theta}
\\
\dot{\phi}
={}&
\Gamma
\frac{|Q|}{\sin\theta}
\sin(\phi+\psi_1),
\label{eq:diss_phi}
\end{align}
where
\begin{equation}
\alpha\equiv |a|^2-|b|^2,
\qquad
Q\equiv ac^*-b^*c
=
|Q|e^{i\psi_1}.
\label{eq:diss_Q}
\end{equation}

A useful structural feature of Eqs.~\eqref{eq:diss_theta} and
\eqref{eq:diss_phi} is that all azimuthal phase dependence is
controlled by the single complex combination $Q$. In particular,
the $S_+$--$S_-$ interference does not generate an independent
second-harmonic phase-pinning term at leading order in the
thermodynamic-limit mean-field dynamics. The same invariant $Q$
therefore controls both the phase locking and the feedback of
the azimuthal phase into the latitude dynamics.

%------------------------------------------------------------
\subsection{Absence of a standard Hopf instability}
\label{subsec:absence_single_linear_hopf}
\label{subsec:szsp_hopf}
%------------------------------------------------------------

Equations~\eqref{eq:diss_theta} and
\eqref{eq:diss_phi} lead to a simple no-Hopf result for a
single linear dissipative perturbation. Denote their right-hand
sides by
\begin{equation}
\dot{\theta}=f(\theta,\phi),
\qquad
\dot{\phi}=g(\theta,\phi).
\end{equation}

We first consider $Q=0$. In this case the angular equations reduce to
\begin{equation}
\dot{\phi}=0,
\qquad
\dot{\theta}=
-\sin\theta
\left(
\Gamma_+-\Gamma_-\cos^2\theta+\Gamma\alpha
\right).
\label{eq:diss_Q_zero}
\end{equation}
The azimuthal direction therefore remains neutral and the local
Jacobian contains only real eigenvalues, one of which is zero.
Consequently, no nonzero-frequency complex-conjugate pair can
cross the imaginary axis.

We next consider $Q\neq0$. At any finite-latitude stationary point,
Eq.~\eqref{eq:diss_phi} requires
\begin{equation}
\sin(\phi^*+\psi_1)=0,
\qquad
\phi^*+\psi_1=n\pi,
\qquad n\in\mathbb Z.
\label{eq:diss_phase_locking}
\end{equation}
At such a phase-locked point, both off-diagonal Jacobian elements
vanish, $f_\phi^*=g_\theta^*=0$. The remaining diagonal elements are
\begin{align}
f_\theta^*
={}&
-\cos\theta^*
\left(
\Gamma_+-\Gamma_-\cos^2\theta^*+\Gamma\alpha
\right)
\nonumber\\
&-2\Gamma_-\sin^2\theta^*\cos\theta^*
+\Gamma|Q|\sin\theta^*(-1)^n,
\label{eq:diss_f_theta}
\\
g_\phi^*
={}&
\Gamma
\frac{|Q|}{\sin\theta^*}
(-1)^n.
\label{eq:diss_g_phi}
\end{align}
The two eigenvalues are therefore
$\lambda_1=f_\theta^*$ and $\lambda_2=g_\phi^*$, and are both real.
The phase-locked stationary point can consequently change stability
only through a zero crossing of a real eigenvalue, rather than through
a finite-frequency complex-conjugate pair.

Combining the $Q=0$ and $Q\neq0$ cases, we conclude that a
single linear dissipative $U(1)$-breaking jump of the form
Eq.~(\ref{eq:linear_jump_general}) cannot generate a standard Hopf
bifurcation within the thermodynamic-limit mean-field dynamics
considered here. A dissipatively induced Hopf bifurcation therefore
requires additional dynamical structure beyond a single linear jump,
such as nonlinear symmetry-breaking dissipation, multiple independent
dissipative channels, or additional coherent or nonlinear couplings
that keep the latitude and phase fluctuations dynamically coupled at
the stationary point.

%============================================================
\section{Discussion and Conclusion}
%============================================================

We have developed a unified dynamical picture of how nonlinear
dissipation and explicit $U(1)$ breaking organize the local bifurcation
structure of driven-dissipative collective spins. The central organizing principle
is the separation between the phase-neutral background and the
perturbation that supplies a phase reference. This separation makes it
possible to distinguish the microscopic mechanism that selects a finite
oscillation amplitude from the subsequent mechanism that locks,
destabilizes, or reorganizes the free phase.

At the background level, the decisive ingredient is nonlinear
dissipation. A single linear covariant jump must belong to one
symmetry-charge sector and therefore generates polar relaxation or
dephasing, but no stable finite-latitude attractor. The nonlinear
covariant channel supplies the missing saturation: it balances gain and
loss while retaining a neutral azimuthal direction, stabilizes the SSO
manifold and drives the north-pole PFP through a supercritical Hopf
bifurcation that creates the stable finite-amplitude SSO branch
at $\Gamma_-=\Gamma_+$. Thus nonlinear dissipation is not
simply a quantitative correction to linear damping; within the minimal
architecture it changes the available attractor topology by creating the
finite-amplitude background on which subsequent phase dynamics takes
place.

The response to coherent $U(1)$ breaking depends qualitatively on both
detuning and background structure. At exact resonance, the relevant
stationary branch reaches a double-zero degeneracy. Its critical
frequency vanishes, and the nearby closed trajectories are not an
isolated attracting cycle selected by a standard Hopf bifurcation.
Finite detuning unfolds this singular onset: a complex-conjugate
eigenvalue pair crosses the imaginary axis with nonzero frequency,
producing a genuine Hopf boundary. Importantly, the trace-zero condition
places this boundary exclusively on the SSO side, where
$\Gamma_->\Gamma_+$. The driven instability therefore retains a clear
memory of the undriven attractor: detuning alone does not create Hopf
criticality when nonlinear dissipation has not first selected the
appropriate oscillatory background. Along the finite-detuning boundary,
the nonlinear Hopf character may be either supercritical or subcritical,
as diagnosed by the first Lyapunov coefficient.

Dissipative symmetry breaking leads to a different constraint. For the
single linear mixed jump, the thermodynamic-limit angular flow contains
only one first-harmonic phase-pinning invariant $Q$; the apparent
$S_+$--$S_-$ interference does not provide an independent second
harmonic. If $Q=0$, the phase direction remains neutral. If $Q\neq0$,
the fixed phase satisfies $\phi^*+\psi_1=n\pi$, precisely where the
linearized latitude and phase fluctuations decouple. The Jacobian is
then diagonal and its two eigenvalues are real. Consequently, a single
linear dissipative $U(1)$-breaking channel can pin the phase but cannot
generate a finite-frequency Hopf bifurcation. This no-Hopf result shows
that phase pinning and oscillatory destabilization are distinct
dynamical capabilities.

These results extend the background-attractor criterion introduced in
our previous work~\cite{Wang2026} by providing its microscopic
dissipative and bifurcation-theoretic basis. Here we identify nonlinear
covariant dissipation as the minimal origin of that background and show
how the symmetry-breaking channel determines its subsequent fate.
Synchronization and boundary time-crystalline oscillations can thus be
viewed as consequences of a more general attractor-selection problem,
in which dissipation first constructs the autonomous oscillatory
manifold and the symmetry-breaking perturbation then locks or
destabilizes its neutral phase.

The present treatment is restricted to thermodynamic-limit mean-field
dynamics and minimal single-channel perturbations. This restriction also
clarifies the next questions. Finite-size fluctuations, the associated
Liouvillian spectral scaling, and more general combinations of nonlinear
or multichannel dissipation remain natural directions for further study.
These extensions do not alter the present design principle: nonlinear
covariant dissipation selects the self-sustained background, while the
structure of the symmetry-breaking channel determines whether the
resulting local response is non-Hopf, double-zero, or genuinely Hopf.

\section*{Acknowledgments}
This work was supported by the National Natural Science Foundation of
China (Grant No. 12505027, No. 12575026, No. 12574297), the Shenzhen
Science and Technology Program under Grant No. JCYJ20250604145221028,
Natural Science Foundation of Top Talent of SZTU under Grant No.
GDRC202527, and the
Natural Science Foundation of Sichuan Province under Grant No.
2025ZNSFSC0058.

\appendix
\section{$U(1)$ covariance of a single linear jump operator}
\label{app:U1_covariant_linear_jump}

This Appendix supplies the symmetry and dynamical arguments used in
Sec.~\ref{subsec:linear_no_sso}. For a single linear jump operator of
the form in Eq.~(\ref{eq:linear_jump_general}), with the overall
prefactor \(\sqrt{\Gamma/S}\) suppressed, \(U(1)\) covariance of the
dissipator forces the jump operator to belong to a single charge sector.
Equivalently, at most one coefficient among \(a,b,c\) can be nonzero.
We then show that the three allowed charge sectors generate only polar
relaxation within this description and therefore cannot stabilize a
finite-latitude SSO background.

The $U(1)$ symmetry is generated by the rotation
\(U(\chi)=e^{-i\chi S^z}\) already introduced in
Eq.~(\ref{eq:U1_rotation}).
Physically, \(U(1)\) covariance means that the rotation operation
commutes with the dissipative evolution: rotating the state before or
after applying the dissipator gives the same result. Accordingly,
A dissipator \(\mathcal D[L]\) is \(U(1)\)-covariant if
for any density matrix $\rho$ and any angle $\chi$,
\begin{equation}
U(\chi)\,\mathcal D[L](\rho)\,U^\dagger(\chi)
=
\mathcal D[L]\!\left(U(\chi)\rho U^\dagger(\chi)\right).
\label{eq:app_U1_covariance_def}
\end{equation}
Using the explicit form
\begin{equation}
\mathcal D[L](\rho)
=
L \rho L^\dagger
-
\frac12\{L^\dagger L,\rho\},
\end{equation}
the rotated dissipator can be written as
\begin{equation}
U(\chi)\,\mathcal D[L](\rho)\,U^\dagger(\chi)
=
\mathcal D[L(\chi)]\!\left(U(\chi)\rho U^\dagger(\chi)\right),
\label{eq:app_D_transform}
\end{equation}
where
\begin{equation}
L(\chi)\equiv U(\chi)L U^\dagger(\chi).
\label{eq:app_Lchi_def}
\end{equation}
Thus the covariance condition
(\ref{eq:app_U1_covariance_def}) is equivalent to
\begin{equation}
\mathcal D[L(\chi)] = \mathcal D[L]
\qquad
\text{for all }\chi.
\label{eq:app_covariance_equiv}
\end{equation}

For a single jump operator, the dissipator is invariant
under a global phase of the jump,
\begin{equation}
\mathcal D[e^{i\eta}L]=\mathcal D[L].
\label{eq:app_global_phase_invariance}
\end{equation}
Consequently, covariance is guaranteed if
\begin{equation}
L(\chi)=e^{i\eta(\chi)}L
\qquad
\text{for some real function }\eta(\chi).
\label{eq:app_phase_covariance_condition}
\end{equation}
For a single nonzero jump operator this phase condition is also
necessary: two jump operators give the same single-channel dissipator
only when they span the same one-dimensional jump space. Hence
\(\mathcal D[L(\chi)] = \mathcal D[L]\) requires \(L(\chi)\) and
\(L\) to differ only by an overall phase.

We now apply Eq.~(\ref{eq:app_phase_covariance_condition}) to the
linear jump sector in Eq.~(\ref{eq:linear_jump_general}). Under the
\(U(1)\) rotation (\ref{eq:U1_rotation}), the collective spin operators
transform as in Eq.~(\ref{eq:U1_spin_transformation}), so
\begin{equation}
L(\chi)
=
a e^{i\chi}S_+
+
b e^{-i\chi}S_-
+
c S_z.
\label{eq:app_Lchi_explicit}
\end{equation}
Requiring \(L(\chi)=e^{i\eta(\chi)}L\) then gives the componentwise
conditions
\begin{equation}
a e^{i\chi}=e^{i\eta(\chi)}a,
\qquad
b e^{-i\chi}=e^{i\eta(\chi)}b,
\qquad
c=e^{i\eta(\chi)}c.
\label{eq:app_component_conditions}
\end{equation}
If $a\neq0$, the first relation implies
\begin{equation}
e^{i\eta(\chi)}=e^{i\chi}.
\label{eq:app_eta_a}
\end{equation}
If $b\neq0$, the second relation implies
\begin{equation}
e^{i\eta(\chi)}=e^{-i\chi}.
\label{eq:app_eta_b}
\end{equation}
If $c\neq0$, the third relation implies
\begin{equation}
e^{i\eta(\chi)}=1.
\label{eq:app_eta_c}
\end{equation}
These three phase assignments are mutually incompatible for generic
\(\chi\). Therefore Eq.~(\ref{eq:app_phase_covariance_condition})
cannot be satisfied when two or more of \(a,b,c\) are simultaneously
nonzero. The only \(U(1)\)-covariant possibilities for a single linear
jump operator are
\(L\propto S_+\), carrying charge \(+1\),
\(L\propto S_-\), carrying charge \(-1\), or
\(L\propto S_z\), carrying charge \(0\).
This proves the symmetry statement used in
Sec.~\ref{subsec:linear_no_sso}: \(L\) must carry a definite
\(U(1)\) charge, so at most one of \(a,b,c\) may be nonzero.

It remains to record the dynamical consequence of the three allowed
charge sectors. Their longitudinal drift has the common form
\begin{equation}
\dot m_z=\Gamma\,\alpha\,(1-m_z^2),
\label{eq:app_mz_linear_generic}
\end{equation}
where the coefficient \(\alpha\) depends on the charge sector:
\(\alpha=|a|^2>0\) for \(L\propto S_+\),
\(\alpha=-|b|^2<0\) for \(L\propto S_-\), and
\(\alpha=0\) for \(L\propto S_z\).
Using the spherical parametrization \(m_z=\cos\theta\) from
Eq.~(\ref{eq:bloch_spherical}), one has
\begin{equation}
\dot m_z=-\sin\theta\,\dot\theta,
\qquad
1-m_z^2=\sin^2\theta.
\label{eq:app_theta_kinematics_linear}
\end{equation}
Substituting \(m_z=\cos\theta\) into
Eq.~(\ref{eq:app_mz_linear_generic}) therefore gives
the polar equations
\begin{equation}
\dot{\theta}
=
\begin{cases}
-\Gamma |a|^2\sin\theta,
&L\propto S_+,\\[4pt]
\Gamma |b|^2\sin\theta,
&L\propto S_-,\\[4pt]
0,
&L\propto S_z,
\end{cases}
\label{eq:app_theta_linear_dynamics}
\end{equation}
already quoted in Eq.~(\ref{eq:theta_linear_dynamics}).
The \(S_+\) and \(S_-\) channels drive the spin monotonically toward a
pole, while the \(S_z\) channel gives no latitude drift. None of these
flows contains a stable fixed point at \(0<\theta^\ast<\pi\). Thus a
single linear \(U(1)\)-covariant jump operator can produce only PFP-type
polar relaxation, or pure dephasing in the \(S_z\) sector, and cannot
generate the finite-amplitude self-saturation required for an SSO
background.

\section{Hopf bifurcation from the PFP background to the SSO background}
\label{app:pfp_sso_hopf}

This Appendix gives the linear and nonlinear ingredients behind the
PFP-to-SSO transition discussed in Sec.~\ref{sec:III_A}. In the
laboratory frame, the transition is a Hopf bifurcation of the
north-pole PFP. The linearized dynamics establishes the
nonzero-frequency eigenvalue crossing, while the transverse-amplitude
equation determines the supercritical character of the bifurcation.

We use Cartesian Bloch variables
\((m_x,m_y,m_z)\), with \(m_\alpha=\langle S_\alpha\rangle/S\). The
background Hamiltonian \(H_0=\omega_0 S_z\) together with the
\(U(1)\)-covariant dissipative channels gives
\begin{align}
\dot m_x &= -\omega_0 m_y - \bigl(\Gamma_+ - \Gamma_- m_z^2\bigr)m_x m_z,
\label{eq:app_mx_bg}\\
\dot m_y &= \phantom{-}\omega_0 m_x - \bigl(\Gamma_+ - \Gamma_- m_z^2\bigr)m_y m_z,
\label{eq:app_my_bg}\\
\dot m_z &= \bigl(\Gamma_+ - \Gamma_- m_z^2\bigr)(1-m_z^2).
\label{eq:app_mz_bg}
\end{align}
We first analyze the north-pole PFP,
\begin{equation}
(m_x,m_y,m_z)=(0,0,1).
\label{eq:app_north_pole}
\end{equation}
We introduce small transverse perturbations
\begin{equation}
m_x=\delta x,\qquad
m_y=\delta y,\qquad
m_z=1+\delta z.
\label{eq:app_perturbations}
\end{equation}
Since \(\delta z=O(\delta x^2+\delta y^2)\), to linear order we may set $m_z\simeq 1$.
Substituting Eq.~\eqref{eq:app_perturbations} into
Eqs.~\eqref{eq:app_mx_bg}--\eqref{eq:app_my_bg}, we obtain the linearized
transverse dynamics
\begin{align}
\delta\dot x &= (\Gamma_- - \Gamma_+)\delta x - \omega_0\,\delta y,
\label{eq:app_dx_lin}\\
\delta\dot y &= \omega_0\,\delta x + (\Gamma_- - \Gamma_+)\delta y.
\label{eq:app_dy_lin}
\end{align}
Therefore the Jacobian in the transverse plane is
\begin{equation}
J_\perp=
\begin{pmatrix}
\Gamma_- - \Gamma_+ & -\omega_0\\
\omega_0 & \Gamma_- - \Gamma_+
\end{pmatrix}.
\label{eq:app_J_perp}
\end{equation}
Its eigenvalues are
\begin{equation}
\lambda_\pm=(\Gamma_- - \Gamma_+)\pm i\omega_0.
\label{eq:app_eigs_hopf}
\end{equation}
Equation~\eqref{eq:app_eigs_hopf} shows that the PFP is linearly stable
for \(\Gamma_-<\Gamma_+\), becomes marginal at
\(\Gamma_-=\Gamma_+\), and loses stability for
\(\Gamma_->\Gamma_+\). At the critical point the eigenvalues are
purely imaginary, \(\lambda_\pm=\pm i\omega_0\), with nonzero
oscillation frequency \(\omega_0\). Moreover,
\begin{equation}
\frac{d}{d(\Gamma_- - \Gamma_+)}\operatorname{Re}\lambda_\pm = 1 \neq 0,
\label{eq:app_transversality}
\end{equation}
so the complex-conjugate pair crosses the imaginary axis transversely.
Therefore, provided that \(\omega_0\neq0\), the north-pole PFP satisfies
the linear spectral conditions for a standard Hopf bifurcation: a
simple complex-conjugate pair crosses the imaginary axis transversely
at a nonzero frequency. This conclusion follows directly from the
spectrum of the transverse Jacobian \(J_\perp\).

\paragraph*{Supercriticality of the background Hopf point}
The linear analysis above establishes a standard Hopf threshold
but does not determine its nonlinear criticality. For the present
PFP-to-SSO transition, the nonlinear type can be determined
directly from the physical transverse amplitude
\begin{equation}
r_{\perp}
\equiv
\sqrt{m_x^2+m_y^2}.
\label{eq:app_r_def}
\end{equation}
We use the notation $r_{\perp}$ here to distinguish this physical
transverse amplitude from the normal-form amplitude $R$
introduced in Sec.~\ref{subsec:trace_determinant_classification}. On the Bloch sphere,
$m_z=\sqrt{1-r_{\perp}^2}$ near the north pole. The corresponding transverse-amplitude
equation obtained from Eqs.~\eqref{eq:app_mx_bg}--\eqref{eq:app_mz_bg} is
\begin{equation}
\dot r_{\perp}
=
\left[
\Gamma_-(1-r_{\perp}^2)-\Gamma_+
\right]
\sqrt{1-r_{\perp}^2}\,
r_{\perp}.
\end{equation}
Writing $\mu=\Gamma_- - \Gamma_+$, its expansion near the Hopf point takes the form
\begin{equation}
\dot r_{\perp}
=
\mu r_{\perp}
-
\Gamma_- r_{\perp}^3
+
O(\mu r_{\perp}^3,r_{\perp}^5).
\label{eq:app_r_normal_form}
\end{equation}
The cubic coefficient is negative. In the sign convention of
Sec.~\ref{subsec:trace_determinant_classification}, this corresponds to the supercritical case. Indeed,
for $\mu>0$, where the north-pole PFP has already lost linear
stability, the transverse-amplitude equation admits the nonzero stationary amplitude
\begin{equation}
r_{\perp}^*
=
\sqrt{
\frac{\Gamma_- - \Gamma_+}{\Gamma_-}
}.
\label{eq:app_r_star}
\end{equation}
Its radial stability follows from
\begin{equation}
\left.
\frac{\partial \dot r_{\perp}}
{\partial r_{\perp}}
\right|_{r_{\perp}=r_{\perp}^*}
=
-2\mu
+
O(\mu^2)
<0,
\qquad
\mu>0,
\label{eq:app_r_radial_stability}
\end{equation}
so the nonzero branch is locally attracting near onset.
Thus a stable small-amplitude oscillatory branch emerges on
the side where the PFP has lost stability, which is precisely
the supercritical Hopf scenario defined in
Sec.~\ref{subsec:trace_determinant_classification}. Moreover,
\begin{equation}
r_{\perp}^*
\propto
\mu^{1/2},
\qquad
\mu\rightarrow0^+,
\label{eq:app_r_square_root_onset}
\end{equation}
showing the expected continuous square-root onset.

In the laboratory frame, this state is an orbitally stable
limit cycle with angular frequency $\omega_0$. In the frame
rotating at $\omega_0$, the same motion is represented by a
continuous finite-latitude SSO manifold of stationary states,
which is transversely attracting but neutral along the
azimuthal phase. Therefore, the PFP-to-SSO background
transition occurs through a supercritical Hopf bifurcation at
$\Gamma_-=\Gamma_+$.

\section{Derivations and eigenvalue diagnostics for the coherent-drive}
\label{app:resonant_threshold_corrected}
This Appendix gives the fixed-point, stability, and threshold
derivations used in Sec.~\ref{sec:coherent_u1_breaking_drive}. At
resonance, where the stationary equations split into two branches, we
determine the physical root multiplicity, branch endpoints, and
Jacobian spectra. For finite detuning, we derive the compact trace and
determinant formulas, Eqs.~(\ref{eq:trace_detuned_compact}) and
(\ref{eq:det_detuned_compact}), and the resulting conditions for a
standard Hopf threshold from the equations
(\ref{eq:theta_epsilon}) and (\ref{eq:phi_epsilon}). The eigenvalue and
phase-portrait diagnostics use representative imbalance parameters
\(\eta=-0.2\) and \(0.2\), while additional resonant fixed-point plots
at \(\eta=0.6\) and \(0.8\) display the crossover at \(\eta=0.6\) and
the coexistence regime. The resonant analysis establishes the
double-zero spectrum, branch multiplicity, and local nonlinear center
structure, without invoking a generic Bogdanov--Takens classification.

\subsection{Resonance}
\label{app:resonance_details}

At \(\Delta=0\), the stationary equations are
\begin{align}
0&=
-\sin\theta^\ast
\bigl(\Gamma_+-\Gamma_-\cos^2\theta^\ast\bigr)
+\frac{\epsilon}{2}\sin\phi^\ast,
\label{eq:app_fp_theta}\\
0&=
\frac{\epsilon}{2}\cot\theta^\ast\cos\phi^\ast.
\label{eq:app_fp_phi}
\end{align}
For \(\epsilon>0\), Eq.~(\ref{eq:app_fp_phi}) gives the two branches
\[
\cos\phi^\ast=0
\qquad\text{or}\qquad
\theta^\ast=\frac{\pi}{2},
\]
in agreement with Eq.~(\ref{eq:res_two_branches_main}). Their
existence and stability are treated separately. The
\(\cos\phi^\ast=0\) branch gives the real-eigenvalue fixed points and
sets the branch multiplicity and endpoints. The equatorial branch,
instead, carries the resonant double-zero onset and the associated
neutral center structure.

\subsubsection{Real-eigenvalue branch: existence, multiplicity, and stability}
\label{app:real_eigenvalue_branch_details}

Writing \(\phi^\ast=s\pi/2\), with
\(s=\sin\phi^\ast=\pm1\), Eq.~(\ref{eq:app_fp_theta}) becomes
\begin{equation}
\sin\theta^\ast
\bigl(\Gamma_+-\Gamma_-\cos^2\theta^\ast\bigr)
=s\,\frac{\epsilon}{2}.
\label{eq:app_theta_fp_reduced}
\end{equation}
With \(x=\cos^2\theta^\ast\in[0,1]\), squaring this equation gives
\begin{equation}
g(x)\equiv
(\Gamma_+-\Gamma_-x)^2(1-x)
=\frac{\epsilon^2}{4}.
\label{eq:app_gx_eq}
\end{equation}
For every root at \(\epsilon>0\), the unsquared equation fixes
\[
s=\operatorname{sgn}(\Gamma_+-\Gamma_-x),
\]
so Eq.~(\ref{eq:app_gx_eq}) introduces no spurious physical root.
Complex roots and real roots outside \(x\in[0,1]\) are discarded.

The shape of \(g(x)\) determines the number of physical roots.
Differentiation gives
\begin{equation}
g'(x)=
-(\Gamma_+-\Gamma_-x)
\bigl(\Gamma_++2\Gamma_--3\Gamma_-x\bigr).
\label{eq:app_gprime}
\end{equation}
The two stationary points are
\begin{equation}
x_1=\frac{\Gamma_+}{\Gamma_-},
\qquad
x_2=\frac{\Gamma_++2\Gamma_-}{3\Gamma_-}.
\label{eq:app_x1_x2_stability}
\end{equation}
Their relevant function values are
\begin{equation}
\begin{aligned}
g(0)&=\Gamma_+^2,~~~g(1)=g(x_1)=0,\\
g(x_2)&=
\frac{4(\Gamma_--\Gamma_+)^3}{27\Gamma_-}
\quad(\Gamma_->\Gamma_+).
\end{aligned}
\label{eq:app_g_values}
\end{equation}
We introduce the two drive scales
\begin{equation}
\epsilon_{x=0}=2\Gamma_+,
\qquad
\epsilon_{x_2}=
\frac{4}{3\sqrt3}
\frac{(\Gamma_--\Gamma_+)^{3/2}}{\sqrt{\Gamma_-}},
\label{eq:app_two_resonant_scales}
\end{equation}
where \(\epsilon_{x_2}\) exists only for \(\Gamma_->\Gamma_+\).
They are the horizontal levels associated with \(g(0)\) and
\(g(x_2)\), respectively.

In the PFP regime, \(\Gamma_+\ge\Gamma_-\) or \(\eta\le0\), both
\(x_1\) and \(x_2\) lie at or beyond \(x=1\), and \(g(x)\) decreases
monotonically on \([0,1]\). Hence there is one physical \(x\) root for
\(0<\epsilon<\epsilon_{x=0}\), the root reaches \(x=0\) at
\(\epsilon=\epsilon_{x=0}\), and no root on this branch
survives for larger drive.

In the SSO regime, \(\Gamma_->\Gamma_+\) or \(\eta>0\), one has
\(0<x_1<x_2<1\). The function decreases from \(g(0)\) to zero at
\(x_1\), rises to the local maximum \(g(x_2)\), and then decreases to
zero at \(x=1\). Comparing the two maxima gives
\begin{equation}
g(0)=g(x_2)
\quad\Longleftrightarrow\quad
\frac{\Gamma_+}{\Gamma_-}=\frac14
\quad\Longleftrightarrow\quad
\eta=\frac35.
\label{eq:app_eta_crossover}
\end{equation}
Thus the PFP regime is governed by a monotonic root geometry, whereas
the SSO regime develops a two-extrema structure and can support branch
coexistence. The resulting physical-root multiplicity is:
\begin{itemize}
\item For \(0<\eta<3/5\), one has
\(\epsilon_{x_2}<\epsilon_{x=0}\). There are three roots for
\(0<\epsilon<\epsilon_{x_2}\), two distinct roots at
\(\epsilon=\epsilon_{x_2}\) (one is double), one root for
\(\epsilon_{x_2}<\epsilon\le\epsilon_{x=0}\), and no root above
\(\epsilon_{x=0}\).
\item For \(\eta=3/5\), the two maxima are equal. At
\(\epsilon=\epsilon_{x_2}=\epsilon_{x=0}\), the physical
solutions are \(x=0\) and the double root \(x=x_2\).
\item For \(3/5<\eta<1\), one has
\(\epsilon_{x=0}<\epsilon_{x_2}\). There are three roots for
\(0<\epsilon\le\epsilon_{x=0}\), two roots for
\(\epsilon_{x=0}<\epsilon<\epsilon_{x_2}\), one double root at
\(\epsilon=\epsilon_{x_2}\), and no root above \(\epsilon_{x_2}\).
\end{itemize}
Consequently, the largest drive for which any
\(\cos\phi^\ast=0\) fixed point exists is
\begin{equation}
\epsilon_{\mathrm{end}}=
\begin{cases}
2\Gamma_+,
&\Gamma_+\ge\Gamma_-/4,\\[6pt]
\dfrac{4}{3\sqrt3}
\dfrac{(\Gamma_--\Gamma_+)^{3/2}}{\sqrt{\Gamma_-}},
&\Gamma_+<\Gamma_-/4.
\end{cases}
\label{eq:app_gmax_piecewise}
\end{equation}
Thus \(\epsilon_{\mathrm{end}}\) is the global existence boundary of the complete
\(\cos\phi^\ast=0\) solution set; it need not coincide with the onset
of the equatorial branch.

At \(\epsilon=0\), Eq.~(\ref{eq:app_gx_eq}) also contains the pole
\(x=1\) and the double zero \(x=x_1\) when \(\eta>0\). These are
special zero-drive limits: the azimuth is undefined at the poles, and
the phase condition is degenerate. They are not counted by the
finite-drive multiplicity statements above.

The root count of Eq.~(\ref{eq:app_gx_eq}) does not by itself give the
fixed-point count. To reconstruct the physical stationary points and their
stability, we distinguish three related notions:
the number of algebraic roots of the cubic equation for \(x\),
the number of roots that lie in the physical interval \(x\in[0,1]\),
and the number of distinct fixed points reconstructed from those
physical roots.
For \(\epsilon>0\), every root \(x_i\in(0,1)\) gives two fixed points,
\begin{align}
\theta_{i,\mathrm N}^\ast
&=\arccos\sqrt{x_i},
\nonumber\\
\theta_{i,\mathrm S}^\ast
&=\pi-\arccos\sqrt{x_i},
\nonumber\\
\phi_i^\ast
&=\frac{\pi}{2}
\operatorname{sgn}(\Gamma_+-\Gamma_-x_i)
\pmod{2\pi}.
\label{eq:app_fixed_points_from_x}
\end{align}
Accordingly, three distinct interior roots represent six fixed
points. At \(x=0\), the northern and southern points coincide and are
counted only once. Thus not every algebraic root corresponds to an
independent physical branch; the multiplicity statements refer to the
branches after this geometric reconstruction.

\begin{figure*}[tp]
\centering
\includegraphics[width=0.85\textwidth]{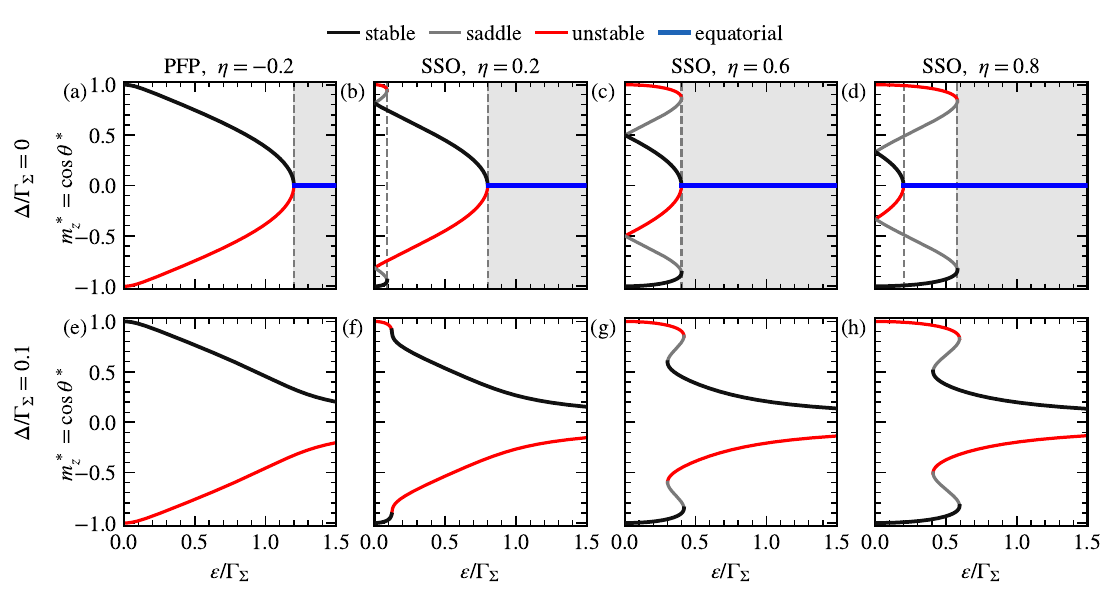}
\caption{
Physical fixed-point branches at \(\Gamma_\Sigma=1\). The four columns
correspond to \(\eta=-0.2,0.2,0.6,\) and \(0.8\), and the two rows to
\(\Delta/\Gamma_\Sigma=0\) and \(0.1\).
Black, gray, red, and blue branches denote stable, saddle, unstable,
and equatorial solutions, respectively. Equatorial branches exist only
at exact resonance. The gray-shaded strong-drive region begins at
\(\epsilon_{\mathrm{end}}\).
}
\label{fig:app_fixed_point_branches}
\end{figure*}

The Jacobian elements are already given in
Eqs.~(\ref{eq:main_j11})--(\ref{eq:main_j22}). On the
\(\cos\phi^\ast=0\) branch the off-diagonal elements vanish, so
\begin{equation}
J_\ast=
\begin{pmatrix}
\lambda_1&0\\
0&\lambda_2
\end{pmatrix},~
\begin{aligned}
\lambda_1&=
\cos\theta^\ast
\bigl(3\Gamma_-x-2\Gamma_--\Gamma_+\bigr),\\
\lambda_2&=
\cos\theta^\ast(\Gamma_-x-\Gamma_+).
\end{aligned}
\label{eq:app_res_jac_diag}
\end{equation}
Using Eq.~(\ref{eq:app_x1_x2_stability}), this becomes
\begin{equation}
\lambda_1=
3\Gamma_-\cos\theta^\ast(x-x_2),~
\lambda_2=
\Gamma_-\cos\theta^\ast(x-x_1).
\label{eq:app_res_eigs_factorized}
\end{equation}
Because the spectrum is real, every hyperbolic fixed point is
classified directly by the two-dimensional Jacobian criteria
\begin{equation}
\begin{aligned}
\det J_\ast>0,\quad\Tr J_\ast<0
&\Longrightarrow\text{stable node},\\
\det J_\ast>0,\quad\Tr J_\ast>0
&\Longrightarrow\text{unstable node},\\
\det J_\ast<0
&\Longrightarrow\text{saddle}.
\end{aligned}
\label{eq:app_real_eigenvalue_criteria}
\end{equation}
These are the criteria used to label the fixed points in
Fig.~\ref{fig:app_fixed_point_branches}; no separate numerical
criterion is introduced.

For \(\eta>0\), the signs in the three intervals are
\begin{equation}
\begin{gathered}
\begin{array}{c|c|c}
\multicolumn{3}{c}{\cos\theta^\ast=+\sqrt{x}\ \text{(northern)}}\\
x\text{ interval}&
(\operatorname{sgn}\lambda_1,\operatorname{sgn}\lambda_2)&
\text{type}\\ \hline
(0,x_1)&(-,-)&\text{stable}\\
(x_1,x_2)&(-,+)&\text{saddle}\\
(x_2,1)&(+,+)&\text{unstable}
\end{array}
\\[6pt]
\begin{array}{c|c|c}
\multicolumn{3}{c}{\cos\theta^\ast=-\sqrt{x}\ \text{(southern)}}\\
x\text{ interval}&
(\operatorname{sgn}\lambda_1,\operatorname{sgn}\lambda_2)&
\text{type}\\ \hline
(0,x_1)&(+,+)&\text{unstable}\\
(x_1,x_2)&(+,-)&\text{saddle}\\
(x_2,1)&(-,-)&\text{stable}
\end{array}.
\end{gathered}
\label{eq:app_north_stability_table}
\end{equation}
Thus, whenever three physical roots are present, the associated six
fixed points comprise two stable nodes, two saddles, and two unstable
nodes. When only one or two roots remain, the same table classifies
them according to their intervals. For \(\eta\le0\), there is at most
one physical root; its northern point is stable and its southern
partner is unstable.

At the interior fold \(x=x_2\),
\begin{equation}
\lambda_1=0,
\qquad
\lambda_2=
\frac{2}{3}\cos\theta^\ast(\Gamma_--\Gamma_+)\ne0.
\label{eq:app_fold_eigenvalues}
\end{equation}
The middle and right roots merge there, confirming that
\(\epsilon=\epsilon_{x_2}\) is a saddle-node bifurcation. At \(x=x_1\),
\(\lambda_2=0\), but \(g(x_1)=0\); this degeneracy belongs to the
zero-drive limit and is not a finite-drive bifurcation. At \(x=0\),
\begin{equation}
\epsilon=\epsilon_{x=0},
\qquad
\lambda_1=\lambda_2=0,
\label{eq:app_equatorial_intersection}
\end{equation}
and the real-eigenvalue branches coalesce at the equator and intersect
the equatorial branch at this degenerate double-zero point.

\begin{figure}[t]
\centering
\includegraphics[width=0.85 \columnwidth]{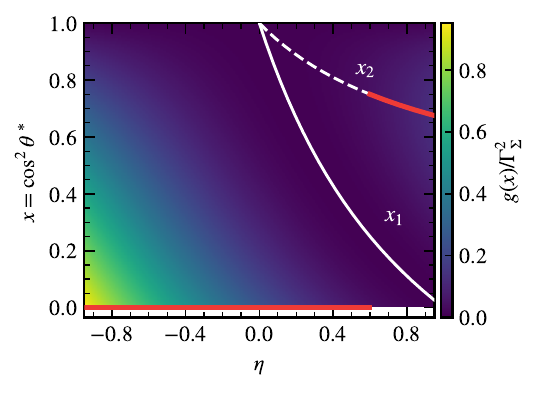}
\caption{
Resonant root geometry for the \(\cos\phi^\ast=0\) branch at
\(\Gamma_\Sigma=\Gamma_++\Gamma_-=1\).
The color map shows \(g(x)\) in the \((\eta,x)\) plane. The solid and
dashed white curves are the stationary points \(x_1\) and \(x_2\),
respectively, and the red curve is the global maximizer of \(g(x)\).
}
\label{fig:app_gx_landscape}
\end{figure}

Figure~\ref{fig:app_gx_landscape} visualizes this root geometry,
including the crossover at \(\eta=3/5\) and the corresponding change
of the resonant existence boundary.

Figure~\ref{fig:app_gx_landscape} therefore provides the geometric
origin of the branch multiplicity discussed above. The physical
reconstruction of those roots is then displayed in
Fig.~\ref{fig:app_fixed_point_branches}. In that figure, the gray
shaded region marks
drive strengths for which the real-eigenvalue
\(\cos\phi^\ast=0\) branch has no physical root. This does not imply
that all fixed points disappear, because the equatorial branch
discussed next can still exist there.
In the first row, panels (a)--(d) correspond to
\(\eta=-0.2,0.2,0.6,\) and \(0.8\) at \(\Delta=0\). Panel (a) lies on
the PFP side and contains only the northern stable node and southern
unstable node on the \(\cos\phi^\ast=0\) branch before they terminate at
the equatorial intersection. By contrast, panels (b)--(d) lie on the
SSO side and develop additional resonant roots, so the
\(\cos\phi^\ast=0\) branch acquires stable, saddle, and unstable
segments before meeting the equatorial branch that continues to larger
drive and carries the double-zero onset.

We next turn from the multiplicity and stability of the
\(\cos\phi^\ast=0\) branch to the Equatorial branch, whose role
is qualitatively different: it carries the
double-zero onset and the neutral oscillatory center structure.

\subsubsection{Equatorial branch}
\label{app:equatorial_branch_details}

The second resonant solution has
\begin{equation}
\theta^\ast=\frac{\pi}{2},
\qquad
\sin\phi^\ast=\frac{2\Gamma_+}{\epsilon},
\label{eq:app_res_equatorial_branch}
\end{equation}
and therefore exists for
\(\epsilon\ge\epsilon_{x=0}=2\Gamma_+\). For
\(\epsilon>\epsilon_{x=0}\), the two phase solutions have
opposite \(\cos\phi^\ast\) but the same spectrum. Their Jacobian and
eigenvalues are
\begin{align}
J_\ast&=
\begin{pmatrix}
0&\dfrac{\epsilon}{2}\cos\phi^\ast\\[4pt]
-\dfrac{\epsilon}{2}\cos\phi^\ast&0
\end{pmatrix},
\label{eq:app_res_equatorial_jac}\\
\lambda_\pm&=
\pm\frac{i}{2}
\sqrt{\epsilon^2-\epsilon_{x=0}^2}.
\label{eq:app_res_equatorial_eigs}
\end{align}
The eigenvalue pair is double-zero at
\(\epsilon=\epsilon_{x=0}\) and becomes purely imaginary above the
threshold. Therefore the linearized flow is
neutrally stable on this branch. Purely imaginary eigenvalues alone do
not establish nonlinear asymptotic stability; the nonlinear analysis
below is needed to distinguish centers from foci or limit cycles. What
is fixed by the linear calculation is that the equatorial branch has a
complex-conjugate neutral spectrum for
\(\epsilon>\epsilon_{x=0}\), while its onset is not a standard
nonzero-frequency Hopf crossing.

For \(\eta\le3/5\), one has
\(\epsilon_{\mathrm{end}}=\epsilon_{x=0}\), so the real-eigenvalue solution
set terminates where the equatorial branch appears. For \(\eta>3/5\),
\(\epsilon_{\mathrm{end}}=\epsilon_{x_2}>\epsilon_{x=0}\), and the two resonant
branches coexist over
\(\epsilon_{x=0}\le\epsilon\le\epsilon_{\mathrm{end}}\).

We denote the resonant onset of the equatorial branch by
\begin{equation}
\epsilon_{\mathrm{DZ}}=\epsilon_{x=0}
=2\Gamma_+ .
\label{eq:app_resonant_dz_threshold}
\end{equation}
The subscript \(\mathrm{DZ}\) emphasizes the double-zero character of
the onset, not a standard Hopf point. To see this distinction
explicitly, write
\(a=(\epsilon/2)\cos\phi^\ast\). Small perturbations of an equatorial
fixed point obey
\begin{equation}
\frac{d}{dt}
\begin{pmatrix}
\delta\theta\\
\delta\phi
\end{pmatrix}
=
\begin{pmatrix}
0&a\\
-a&0
\end{pmatrix}
\begin{pmatrix}
\delta\theta\\
\delta\phi
\end{pmatrix},
\qquad
\delta\ddot{\theta}+a^2\delta\theta=0.
\label{eq:app_resonant_linear_oscillation}
\end{equation}
Thus the imaginary part of the eigenvalue pair is the angular
frequency of the reciprocal oscillation between the polar and
azimuthal perturbations:
\begin{equation}
\omega_{\mathrm{DZ}}(\epsilon)
=
|a|
=
\frac{1}{2}
\sqrt{\epsilon^2-
\epsilon_{\mathrm{DZ}}^2},
\qquad
\epsilon\ge\epsilon_{\mathrm{DZ}},
\label{eq:app_resonant_hopf_frequency}
\end{equation}
where Eq.~(\ref{eq:app_res_equatorial_branch}) was used to eliminate
\(\phi^\ast\). Close to onset,
\begin{equation}
\omega_{\mathrm{DZ}}
\simeq
\sqrt{
\frac{\epsilon_{\mathrm{DZ}}}{2}
\bigl(\epsilon-\epsilon_{\mathrm{DZ}}\bigr)},
\label{eq:app_resonant_soft_mode}
\end{equation}
so the mode softens and its period diverges at the threshold.
Nevertheless,
\(\lambda_+=\lambda_-=0\) at onset, and
\(\operatorname{Re}\lambda_\pm=0\) along the entire equatorial branch.
There is therefore neither a finite critical frequency nor a
transverse crossing of the imaginary axis. In the strict local
bifurcation sense this is a reversible double-zero degeneracy, not a standard
Hopf bifurcation.

Because the reduced resonant flow is already
two-dimensional, this local classification can be obtained directly
from the full flow without an additional center-manifold reduction.
The nonlinear dynamics can be resolved by expanding about this
reversible double-zero point. Define
\begin{equation}
u=\cos\theta,
\qquad
q=\phi-\frac{\pi}{2},
\qquad
\mu=\frac{\epsilon-\epsilon_{\mathrm{DZ}}}{2}.
\label{eq:app_resonant_local_variables}
\end{equation}
In these variables the full resonant flow is
\begin{align}
\dot u
&=
(1-u^2)(\Gamma_+-\Gamma_-u^2)
-\frac{\epsilon}{2}\sqrt{1-u^2}\cos q,
\nonumber\\
\dot q
&=
-\frac{\epsilon}{2}
\frac{u}{\sqrt{1-u^2}}\sin q .
\label{eq:app_resonant_uq_flow}
\end{align}
To the lowest nonvanishing order, the resonant equations become
\begin{align}
\dot u
&=
-\mu
-\left(\Gamma_-+\frac{\Gamma_+}{2}\right)u^2
+\frac{\Gamma_+}{2}q^2
+O_3,
\nonumber\\
\dot q
&=
-\Gamma_+u q
+O_3,
\label{eq:app_resonant_double_zero_normal_form}
\end{align}
where \(O_3\) denotes terms of weighted order higher than two under
the double-zero scaling
\(\mu=O(u^2)=O(q^2)\).
For \(\mu<0\), the local fixed points are
\begin{equation}
q^\ast=0,
\qquad
u_\pm^\ast
=
\pm
\sqrt{
\frac{-\mu}{\Gamma_-+\Gamma_+/2}}
+O(|\mu|).
\label{eq:app_resonant_nodes_below}
\end{equation}
Linearization of Eq.~(\ref{eq:app_resonant_double_zero_normal_form})
shows that \(u_+^\ast\) is a stable node and \(u_-^\ast\) is an
unstable node. They are the northern and southern members of the
real-eigenvalue branch approaching \(x=0\).

For \(\mu>0\), these two nodes are replaced locally by the two
equatorial fixed points
\begin{equation}
u^\ast=0,
\qquad
q_\pm^\ast
=
\pm\sqrt{\frac{2\mu}{\Gamma_+}}
+O(\mu^{3/2}).
\label{eq:app_resonant_centers_above}
\end{equation}
Their linear eigenvalues are
\begin{equation}
\lambda_\pm
=
\pm i\sqrt{2\Gamma_+\mu}
+O(\mu),
\label{eq:app_resonant_centers_eigenvalues}
\end{equation}
which agrees with the soft-mode expansion
Eq.~(\ref{eq:app_resonant_soft_mode}).

The classification can be extended beyond linear order. In the
\((u,\phi)\) variables, the resonant vector field has the
reversibility
\begin{equation}
(u,\phi,t)\longmapsto(-u,\phi,-t).
\label{eq:app_resonant_reversibility}
\end{equation}
Each equatorial fixed point lies on the fixed set \(u=0\) of this
time-reversal involution. An attracting or repelling focus would
violate Eq.~(\ref{eq:app_resonant_reversibility}), because time
reversal interchanges attraction and repulsion while leaving the same
fixed point unchanged. Consequently, the two equatorial fixed points
are nonlinear centers in the ideal thermodynamic-limit dynamics, each
surrounded locally by a continuous family of neutrally stable closed
orbits. The reversible double-zero transition therefore converts a
stable and an unstable node into two nonlinear centers; it does not
create an isolated attracting limit cycle. Detuning, noise, or other
symmetry-breaking perturbations can destroy this neutral family and
select attracting or repelling motion. The resonant onset is therefore
a reversible double-zero degeneracy rather than a Hopf bifurcation.

For \(\eta\le3/5\), the resonant oscillatory onset coincides with
\(\epsilon_{\mathrm{end}}\). For \(\eta>3/5\), instead,
\(\epsilon_{\mathrm{DZ}}=\epsilon_{x=0}
<\epsilon_{x_2}=\epsilon_{\mathrm{end}}\): the equatorial branch appears first,
while the remaining real-eigenvalue fixed points terminate later at
the saddle-node threshold \(\epsilon_{x_2}\). Hence only the former is
the resonant reversible double-zero onset; the latter is an independent
real-eigenvalue bifurcation.

The corresponding nonlinear trajectories are shown in the resonant
columns of Fig.~\ref{fig:resonant_detuned_phase_portraits}. The plotted
curves are obtained by integrating the full resonant
equations, rather than the truncated normal form
Eq.~(\ref{eq:app_resonant_double_zero_normal_form}).

\subsection{Finite detuning}
\label{app:finite_detuning_details}

For \(\Delta\ne0\), the stationary equations give
\begin{align}
\cos\phi^\ast
&=-\frac{2\Delta}{\epsilon}\tan\theta^\ast,
\label{eq:app_phi_detuned}\\
\sin\phi^\ast
&=\frac{2\sin\theta^\ast}{\epsilon}
\bigl(\Gamma_+-\Gamma_-\cos^2\theta^\ast\bigr).
\label{eq:app_sinphi_detuned}
\end{align}
Combining their squares reproduces
Eq.~(\ref{eq:detuned_fixed_point_implicit}). In contrast to exact
resonance, the equatorial solution is no longer an independent branch:
Eq.~(\ref{eq:app_phi_detuned}) is singular at
\(\cos\theta^\ast=0\) unless \(\Delta=0\). The remaining detuned fixed
points are therefore classified by the Jacobian in
Eqs.~(\ref{eq:main_j11})--(\ref{eq:main_j22}) and by the local
two-dimensional stability criteria
\begin{align}
\det J_\ast<0
&\Rightarrow \text{saddle},
\nonumber\\
\det J_\ast>0,\ \Tr J_\ast<0
&\Rightarrow \text{stable},
\label{eq:app_detuned_stability_criteria}\\
\det J_\ast>0,\ \Tr J_\ast>0
&\Rightarrow \text{unstable},
\nonumber\\
\det J_\ast>0,\ \Tr J_\ast=0
&\Rightarrow \text{standard Hopf candidate}.
\nonumber
\end{align}
These criteria underlie the classifications shown in the detuned
fixed-point and eigenvalue plots. We now identify the detuned
trace-zero branch explicitly and determine when its complex pair
crosses the imaginary axis with nonzero frequency.

Using Eq.~(\ref{eq:app_sinphi_detuned}), the lower diagonal element at
a stationary point becomes
\begin{equation}
J_{22}^\ast=
-\cos\theta^\ast
\bigl(\Gamma_+-\Gamma_-\cos^2\theta^\ast\bigr).
\label{eq:app_detuned_j22_reduced}
\end{equation}
Adding \(J_{11}^\ast\) then gives
\begin{align}
\Tr J_\ast
&=
2\cos\theta^\ast
\bigl(2\Gamma_-\cos^2\theta^\ast-\Gamma_--\Gamma_+\bigr)
\nonumber\\
&=
-2\cos\theta^\ast
\bigl(\Gamma_+-\Gamma_-\cos2\theta^\ast\bigr),
\label{eq:app_trace_detuned_derivation}
\end{align}
which is Eq.~(\ref{eq:trace_detuned_compact}).

Similarly, Eq.~(\ref{eq:app_phi_detuned}) gives
\begin{equation}
J_{12}^\ast=-\Delta\tan\theta^\ast,
\qquad
J_{21}^\ast=\frac{\Delta}{\sin\theta^\ast\cos\theta^\ast},
\label{eq:app_detuned_offdiagonal}
\end{equation}
and hence
\begin{align}
\det J_\ast
={}&
\cos^2\theta^\ast
\bigl(\Gamma_+-\Gamma_-\cos^2\theta^\ast\bigr)
\nonumber\\
&\times
\bigl(\Gamma_++2\Gamma_-
-3\Gamma_-\cos^2\theta^\ast\bigr)
+\frac{\Delta^2}{\cos^2\theta^\ast}.
\label{eq:app_det_detuned_derivation}
\end{align}
This reproduces Eq.~(\ref{eq:det_detuned_compact}).

A standard Hopf point requires
\(\Tr J_\ast=0\), \(\det J_\ast>0\), and a nonzero crossing speed.
The trace equation has two formal factors. The choice
\(\cos\theta^\ast=0\) is impossible when \(\Delta\ne0\), because the
stationary phase equation would give \(2\Delta=0\). The remaining
branch is
\begin{equation}
\cos2\theta^\ast=\frac{\Gamma_+}{\Gamma_-},
\qquad
\tan^2\theta^\ast=
\frac{\Gamma_--\Gamma_+}{\Gamma_-+\Gamma_+},
\label{eq:app_detuned_trace_branch}
\end{equation}
which requires \(\Gamma_->\Gamma_+\), or \(\eta>0\). Thus no
standard Hopf point at finite detuning exists in the PFP regime represented by
\(\eta=-0.2\).
When this condition is satisfied, the Hopf candidate has
\begin{equation}
\cos\theta_{\mathrm H}^\ast
=
\sigma
\sqrt{\frac{\Gamma_-+\Gamma_+}{2\Gamma_-}},
\qquad
\sigma=\pm1,
\label{eq:app_detuned_theta_reconstruction}
\end{equation}
and its phase is fixed, modulo \(2\pi\), by
\begin{equation}
\begin{aligned}
\cos\phi_{\mathrm H}^\ast
&=
-\frac{2\Delta}{\epsilon_{\mathrm H}}
\tan\theta_{\mathrm H}^\ast,
\\
\sin\phi_{\mathrm H}^\ast
&=
-\frac{\Gamma_- - \Gamma_+}{\epsilon_{\mathrm H}}
\sqrt{\frac{\Gamma_- - \Gamma_+}{2\Gamma_-}}.
\end{aligned}
\label{eq:app_detuned_phi_reconstruction}
\end{equation}
Equations~(\ref{eq:app_detuned_theta_reconstruction}) and
(\ref{eq:app_detuned_phi_reconstruction}) reconstruct the two
north--south related stationary points on the trace-zero branch; these
are the fixed points whose eigenvalue pair crosses the imaginary axis.

On the trace-zero branch,
\begin{equation}
\det J_\ast=
-\frac{(\Gamma_--\Gamma_+)^2(\Gamma_-+\Gamma_+)}{8\Gamma_-}
+\frac{2\Gamma_-\Delta^2}{\Gamma_-+\Gamma_+}.
\label{eq:app_detuned_det_trace_zero}
\end{equation}
Therefore the complex pair has nonzero frequency only when
\begin{equation}
\Delta^2>
\frac{(\Gamma_--\Gamma_+)^2(\Gamma_-+\Gamma_+)^2}
{16\Gamma_-^2}.
\label{eq:app_detuned_delta_bound}
\end{equation}
At equality, \(\det J_\ast=0\), so the frequency vanishes and the point
is not a standard Hopf bifurcation. Substituting the trace-zero branch
into the stationary amplitude condition reproduces
Eq.~(\ref{eq:detuned_epsilon_bound}),
\begin{equation*}
\epsilon_{\mathrm H}^2(\Delta)=
\frac{(\Gamma_--\Gamma_+)^3}{2\Gamma_-}
+\frac{4(\Gamma_--\Gamma_+)}
{\Gamma_-+\Gamma_+}\Delta^2.
\end{equation*}
Together with the detuning bound in
Eq.~(\ref{eq:app_detuned_delta_bound}), this expression gives the
finite-detuning Hopf boundaries summarized in
Fig.~\ref{fig:detuning_phase_boundary}.
The corresponding Hopf frequency is
\begin{equation}
\begin{aligned}
\omega_{\mathrm H}(\Delta)
&=
\sqrt{\det J_\ast\big|_{\Tr J_\ast=0}}
\\
&=
\left[
-\frac{(\Gamma_--\Gamma_+)^2(\Gamma_-+\Gamma_+)}{8\Gamma_-}
+\frac{2\Gamma_-\Delta^2}{\Gamma_-+\Gamma_+}
\right]^{1/2}.
\end{aligned}
\label{eq:app_detuned_hopf_frequency}
\end{equation}
Finally,
\begin{equation}
\left.
\frac{d\,\Tr J_\ast}{d\theta^\ast}
\right|_{\Tr J_\ast=0}
=-4\Gamma_-\cos\theta^\ast\sin2\theta^\ast\ne0.
\label{eq:app_detuned_trace_slope}
\end{equation}
For the stationary curve parametrized by \(\theta^\ast\),
\begin{equation}
\left.
\frac{d\epsilon^2}{d\theta^\ast}
\right|_{\Tr J_\ast=0}
=
-\sin2\theta^\ast
\left[
(\Gamma_--\Gamma_+)^2
-\frac{16\Gamma_-^2\Delta^2}{(\Gamma_-+\Gamma_+)^2}
\right].
\label{eq:app_detuned_drive_slope}
\end{equation}
Thus the crossing is transverse with respect to the experimentally
tuned drive whenever Eq.~(\ref{eq:app_detuned_drive_slope}) is
nonzero. If this derivative also vanishes, the point is an additional
codimension-two degeneracy rather than the generic Hopf threshold
considered here.

To determine whether the detuned Hopf point is supercritical or
subcritical, we evaluate the first Lyapunov coefficient \(l_1\)
numerically on the true Hopf locus. Writing the translated local flow in
the standard form
\begin{equation}
\dot{\mathbf x}
=
J_{\mathrm H}\mathbf x
+
\frac12 B(\mathbf x,\mathbf x)
+
\frac16 C(\mathbf x,\mathbf x,\mathbf x)
+
O(|\mathbf x|^4),
\label{eq:app_detuned_l1_expansion}
\end{equation}
and normalizing the right and left Hopf eigenvectors
\(q\) and \(p\) by
\(\langle p,q\rangle=1\), we compute \(l_1\) from the standard Hopf
normal-form formula~\cite{Kuznetsov2004}
\begin{equation}
\begin{aligned}
l_1
=
\frac{1}{2\omega_{\mathrm H}}
\Re\Big\langle p,\,
&C(q,q,\bar q)
-2B\!\left(q,J_{\mathrm H}^{-1}B(q,\bar q)\right)
\\
&+
B\!\left(\bar q,
\left(2i\omega_{\mathrm H}I-J_{\mathrm H}\right)^{-1}
B(q,q)\right)
\Big\rangle .
\end{aligned}
\label{eq:app_detuned_l1_formula}
\end{equation}
Here \(\omega_{\mathrm H}\) is given by
Eq.~(\ref{eq:app_detuned_hopf_frequency}).
Figure~\ref{fig:detuned_hopf_dynamics_l1} combines representative
time traces for supercritical and subcritical detuned SSO Hopf points
with the corresponding \(l_1\) scan on the detuned Hopf locus. The
\(l_1\) panel uses the same \(|\Delta|/\Gamma_\Sigma\)-based
presentation as the detuned threshold plot in
Fig.~\ref{fig:detuning_phase_boundary}(b).

The scan shows that the finite-detuning coherent-drive boundary is not
uniform at the nonlinear level: the \(\eta=0.2\) and \(0.6\) branches
remain supercritical throughout the plotted range, whereas the
\(\eta=0.8\) branch crosses from supercritical to subcritical near
\(|\Delta|/\Gamma_\Sigma\approx0.39\). The accompanying time traces
make this distinction dynamical: the supercritical case develops the
small oscillation continuously near threshold, while the subcritical
case requires the finite-amplitude basin of the outer stable cycle.
This closes the nonlinear classification of the representative detuned
Hopf points before we turn to the corresponding eigenvalue plots.

Figure~\ref{fig:app_detuned_eigenvalue_crossing} illustrates the
eigenvalue consequence of this derivation. In
Fig.~\ref{fig:app_detuned_eigenvalue_crossing}, the first row gives the
resonant reference, while the second row shows the finite-detuning case
\(\Delta/\Gamma_\Sigma=0.5\). The four columns correspond to
\(\eta=-0.2,0.2,0.6,\) and \(0.8\), respectively. For the PFP
representative \(\eta=-0.2\), corresponding to
\((\Gamma_+,\Gamma_-)=(0.6,0.4)\), the detuned spectrum never develops a
trace-zero crossing, in agreement with the absence of a standard Hopf threshold at finite detuning.
For the SSO representative \(\eta=0.2\), corresponding to
\((\Gamma_+,\Gamma_-)=(0.4,0.6)\), the resonant onset is a reversible double-zero
degeneracy, whereas at finite detuning the real part of the complex pair
crosses zero at the threshold obtained from
Eq.~(\ref{eq:detuned_epsilon_bound}) while the imaginary part remains
nonzero. The associated phase portraits, shown in the finite-detuning
columns of Fig.~\ref{fig:resonant_detuned_phase_portraits}, display
monotonic relaxation in the PFP case and the reorganization from a
stable fixed point to an oscillatory attractor in the SSO case.
Figure~\ref{fig:detuned_hopf_dynamics_l1} refines this statement at the
nonlinear level: among the three representative detuned SSO points
shown in Fig.~\ref{fig:app_detuned_eigenvalue_crossing}(f)--(h), the
\(\eta=0.2\) and \(0.6\) cases are supercritical, whereas the
\(\eta=0.8\) case is subcritical.

\section{Mean-field angular dynamics of a general linear dissipative jump}
\label{app:mixed_jump_angular_flow}

In this Appendix, we derive the angular contribution generated by the
general linear collective jump operator
\begin{equation}
L=\sqrt{\frac{\Gamma}{S}}\,K,
\qquad
K=aS_{+}+bS_{-}+cS_{z},
\label{app:eq:general_jump}
\end{equation}
where $a$, $b$, and $c$ are in general complex coefficients. Using
$S_{\pm}=S_x\pm iS_y$, the operator $K$ can be written as
\begin{equation}
K=\boldsymbol{\ell}\cdot\mathbf S,
\qquad
\boldsymbol{\ell}
=\bigl(a+b,\;i(a-b),\;c\bigr),
\label{app:eq:ell_definition}
\end{equation}
with $\mathbf S=(S_x,S_y,S_z)$.

\subsection{Classical Bloch-vector drift}

For an arbitrary operator $O$, the adjoint Lindblad dissipator is
\begin{equation}
\mathcal D^\dagger[L]O
=\frac{1}{2}\left(L^\dagger[O,L]+[L^\dagger,O]L\right).
\label{app:eq:adjoint_dissipator}
\end{equation}
Introducing the normalized collective magnetization
$m_i=\langle S_i\rangle/S$, the jump contribution is
\begin{align}
\left.\dot m_i\right|_L
&=\frac{1}{S}\left\langle\mathcal D^\dagger[L]S_i\right\rangle
\nonumber\\
&=\frac{\Gamma}{2S^2}
\left\langle K^\dagger[S_i,K]+[K^\dagger,S_i]K\right\rangle.
\label{app:eq:mi_start}
\end{align}
The collective-spin commutation relations give
\begin{equation}
[S_i,K]=i\epsilon_{ijk}\ell_jS_k,
\qquad
[K^\dagger,S_i]=-i\epsilon_{ijk}\ell_j^*S_k.
\label{app:eq:commutators}
\end{equation}
Thus
\begin{equation}
\left.\dot m_i\right|_L
=\frac{i\Gamma}{2S^2}\epsilon_{ijk}
\left[\ell_j\langle K^\dagger S_k\rangle
-\ell_j^*\langle S_kK\rangle\right].
\label{app:eq:mi_before_factorization}
\end{equation}
In the thermodynamic-limit mean-field approximation,
\begin{equation}
\langle S_iS_j\rangle=S^2m_im_j+O(S),
\label{app:eq:factorization}
\end{equation}
and hence
\begin{align}
\langle K^\dagger S_k\rangle
&=S^2(\boldsymbol{\ell}^*\cdot\mathbf m)m_k+O(S),\\
\langle S_kK\rangle
&=S^2m_k(\boldsymbol{\ell}\cdot\mathbf m)+O(S).
\end{align}
Defining $z\equiv\boldsymbol{\ell}\cdot\mathbf m$ and using
$z^*=\boldsymbol{\ell}^*\cdot\mathbf m$ because $\mathbf m$ is real,
we obtain
\begin{align}
\left.\dot m_i\right|_L
&=\frac{i\Gamma}{2}\epsilon_{ijk}m_k
\left(z^*\ell_j-z\ell_j^*\right)+O(S^{-1})
\nonumber\\
&=-\Gamma\epsilon_{ijk}\operatorname{Im}(z^*\ell_j)m_k
+O(S^{-1}).
\end{align}
In vector notation,
\begin{equation}
\left.\dot{\mathbf m}\right|_L
=-\Gamma\operatorname{Im}
\left[(\boldsymbol{\ell}^*\cdot\mathbf m)\boldsymbol{\ell}\right]
\times\mathbf m+O(S^{-1}).
\label{app:eq:classical_vector_drift}
\end{equation}

\subsection{Projection onto spherical coordinates}

We parametrize the Bloch sphere by
\begin{equation}
\mathbf m=\mathbf e_r
=(\sin\theta\cos\phi,\,\sin\theta\sin\phi,\,\cos\theta).
\label{app:eq:spherical_m}
\end{equation}
The local unit vectors are
\begin{align}
\mathbf e_\theta
&=(\cos\theta\cos\phi,\,\cos\theta\sin\phi,\,-\sin\theta),\\
\mathbf e_\phi&=(-\sin\phi,\cos\phi,0),
\end{align}
so that
\begin{equation}
\dot{\mathbf m}=\dot\theta\,\mathbf e_\theta
+\sin\theta\,\dot\phi\,\mathbf e_\phi.
\label{app:eq:m_dot_spherical}
\end{equation}
For compactness, let $s\equiv\sin\theta$ and $u\equiv\cos\theta$.
The projections of $\boldsymbol{\ell}$ onto the spherical basis are
\begin{align}
z\equiv\boldsymbol{\ell}\cdot\mathbf m
&=s\left(ae^{i\phi}+be^{-i\phi}\right)+cu,
\label{app:eq:z_spherical}\\
\ell_\phi\equiv\boldsymbol{\ell}\cdot\mathbf e_\phi
&=i\left(ae^{i\phi}-be^{-i\phi}\right),
\label{app:eq:ell_phi}\\
\ell_\theta\equiv\boldsymbol{\ell}\cdot\mathbf e_\theta
&=u\left(ae^{i\phi}+be^{-i\phi}\right)-cs.
\label{app:eq:ell_theta}
\end{align}
Writing
$\mathbf A\equiv\operatorname{Im}(z^*\boldsymbol{\ell})
=A_r\mathbf e_r+A_\theta\mathbf e_\theta+A_\phi\mathbf e_\phi$,
we have
$\mathbf A\times\mathbf e_r=A_\phi\mathbf e_\theta
-A_\theta\mathbf e_\phi$. Comparing Eqs.~\eqref{app:eq:classical_vector_drift}
and \eqref{app:eq:m_dot_spherical} therefore gives
\begin{align}
\left.\dot\theta\right|_L
&=-\Gamma\operatorname{Im}(z^*\ell_\phi),
\label{app:eq:theta_projection}\\
\sin\theta\left.\dot\phi\right|_L
&=\Gamma\operatorname{Im}(z^*\ell_\theta).
\label{app:eq:phi_projection}
\end{align}

\subsection{Evaluation for the general jump}

For the latitude equation, Eqs.~\eqref{app:eq:z_spherical} and
\eqref{app:eq:ell_phi} give
\begin{align}
z^*\ell_\phi
={}&is\left[|a|^2-|b|^2-a^*be^{-2i\phi}+ab^*e^{2i\phi}\right]
\nonumber\\
&+iu\left[ac^*e^{i\phi}-bc^*e^{-i\phi}\right].
\label{app:eq:z_ellphi_expanded}
\end{align}
The two $a$--$b$ interference terms obey
\begin{equation}
-a^*be^{-2i\phi}+ab^*e^{2i\phi}=X-X^*,
\qquad X=ab^*e^{2i\phi},
\end{equation}
and are therefore purely imaginary. Their product with $i$ is real and
does not contribute to $\operatorname{Im}(z^*\ell_\phi)$. Hence
\begin{equation}
\operatorname{Im}(z^*\ell_\phi)
=s\left(|a|^2-|b|^2\right)
+u\operatorname{Re}\left[ac^*e^{i\phi}-bc^*e^{-i\phi}\right].
\label{app:eq:theta_intermediate}
\end{equation}
Using
$\operatorname{Re}(bc^*e^{-i\phi})
=\operatorname{Re}(b^*ce^{i\phi})$, define
\begin{equation}
\alpha\equiv|a|^2-|b|^2,
\qquad
Q\equiv ac^*-b^*c=|Q|e^{i\psi_1}.
\label{app:eq:alpha_Q}
\end{equation}
It follows that
\begin{equation}
\operatorname{Im}(z^*\ell_\phi)
=\alpha\sin\theta+|Q|\cos\theta\cos(\phi+\psi_1),
\end{equation}
and therefore
\begin{equation}
\left.\dot\theta\right|_L
=-\Gamma\left[
\alpha\sin\theta+|Q|\cos\theta\cos(\phi+\psi_1)
\right].
\label{app:eq:theta_jump}
\end{equation}

For the azimuthal equation, Eqs.~\eqref{app:eq:z_spherical} and
\eqref{app:eq:ell_theta} yield
\begin{align}
z^*\ell_\theta
={}&su\left[|a|^2+|b|^2+a^*be^{-2i\phi}+ab^*e^{2i\phi}\right]
\nonumber\\
&-cs^2\left(a^*e^{-i\phi}+b^*e^{i\phi}\right)
\nonumber\\
&+c^*u^2\left(ae^{i\phi}+be^{-i\phi}\right)-|c|^2su.
\label{app:eq:z_elltheta_expanded}
\end{align}
The first and last terms are real. For the remaining terms,
\begin{align}
\operatorname{Im}\left[u^2ac^*e^{i\phi}-s^2a^*ce^{-i\phi}\right]
&=\operatorname{Im}\left(ac^*e^{i\phi}\right),\\
\operatorname{Im}\left[u^2bc^*e^{-i\phi}-s^2b^*ce^{i\phi}\right]
&=-\operatorname{Im}\left(b^*ce^{i\phi}\right),
\end{align}
where $u^2+s^2=1$ was used. Consequently,
\begin{align}
\operatorname{Im}(z^*\ell_\theta)
&=\operatorname{Im}\left[(ac^*-b^*c)e^{i\phi}\right]
\nonumber\\
&=|Q|\sin(\phi+\psi_1).
\label{app:eq:phi_Q}
\end{align}
Thus
\begin{equation}
\left.\dot\phi\right|_L
=\Gamma\frac{|Q|}{\sin\theta}\sin(\phi+\psi_1).
\label{app:eq:phi_jump}
\end{equation}
Equations~\eqref{app:eq:theta_jump} and \eqref{app:eq:phi_jump}
show explicitly that the $S_+$--$S_-$ interference produces no
independent second-harmonic term: all phase dependence is governed by
the single invariant $Q$.

Adding the rotating-frame SSO background, which is azimuthally neutral,
gives
\begin{align}
\dot\theta
={}&-\sin\theta\left(\Gamma_+-\Gamma_-\cos^2\theta\right)
\nonumber\\
&-\Gamma\left[
\alpha\sin\theta+|Q|\cos\theta\cos(\phi+\psi_1)
\right],
\label{app:eq:angular_theta_final}\\
\dot\phi
={}&\Gamma\frac{|Q|}{\sin\theta}\sin(\phi+\psi_1).
\label{app:eq:angular_phi_final}
\end{align}
These are Eqs.~\eqref{eq:diss_theta} and \eqref{eq:diss_phi} of the
main text.

\subsection{Linearization at the phase-locked fixed points}

Let the right-hand sides of Eqs.~\eqref{app:eq:angular_theta_final}
and \eqref{app:eq:angular_phi_final} be $f(\theta,\phi)$ and
$g(\theta,\phi)$, respectively. If $Q=0$, then $g=0$: the azimuthal
direction is neutral and the Jacobian eigenvalues are real, with one
identically zero eigenvalue.

For $Q\neq0$, any finite-latitude stationary point must satisfy
\begin{equation}
\sin(\phi^*+\psi_1)=0,
\qquad
\phi^*+\psi_1=n\pi.
\label{app:eq:phase_locked_condition}
\end{equation}
At such a point, the off-diagonal entries vanish:
\begin{align}
f_\phi^*
&=\Gamma|Q|\cos\theta^*\sin(\phi^*+\psi_1)=0,\\
g_\theta^*
&=-\Gamma|Q|\frac{\cos\theta^*}{\sin^2\theta^*}
\sin(\phi^*+\psi_1)=0.
\end{align}
The diagonal entries are
\begin{align}
f_\theta^*
={}&-\cos\theta^*
\left(\Gamma_+-\Gamma_-\cos^2\theta^*+\Gamma\alpha\right)
\nonumber\\
&-2\Gamma_-\sin^2\theta^*\cos\theta^*
+\Gamma|Q|\sin\theta^*(-1)^n,
\label{app:eq:locked_f_theta}\\
g_\phi^*
={}&\Gamma\frac{|Q|}{\sin\theta^*}(-1)^n.
\label{app:eq:locked_g_phi}
\end{align}
Therefore
\begin{equation}
J^*=\begin{pmatrix}f_\theta^*&0\\0&g_\phi^*\end{pmatrix},
\qquad
\lambda_1=f_\theta^*,\quad\lambda_2=g_\phi^*,
\end{equation}
and both eigenvalues are real. Combining the $Q=0$ and $Q\neq0$
cases proves that a single general linear dissipative jump cannot
generate a standard Hopf instability within the thermodynamic-limit
mean-field dynamics considered here.

\bibliographystyle{elsarticle-num}
\bibliography{QSBTC_Refs}

@book{strogatz2024nonlinear,
  title={Nonlinear dynamics and chaos: with applications to physics, biology, chemistry, and engineering},
  author={Strogatz, Steven H},
  year={2024},
  publisher={Chapman and Hall/CRC}
}

@book{pikovsky2001synchronization,
  title={{Synchronization: A Universal Concept in Nonlinear Sciences}},
  author={Pikovsky, A. and Rosenblum, M. and Kurths, J. and Cambridge University Press and Cvitanovic, P. and Moss, F. and Swinney, H.},
  isbn={9780521592857},
  lccn={2003283137},
  series={Cambridge Nonlinear Science Series},
  url={https://books.google.com.hk/books?id=FuIv845q3QUC},
  year={2001},
  publisher={Cambridge University Press}
}

@article{PRL2018Spin1,
  title = {Synchronizing the Smallest Possible System},
  author = {Roulet, Alexandre and Bruder, Christoph},
  journal = {Phys. Rev. Lett.},
  volume = {121},
  issue = {5},
  pages = {053601},
  numpages = {5},
  year = {2018},
  month = {Jul},
  publisher = {American Physical Society},
  doi = {10.1103/PhysRevLett.121.053601},
  url = {https://link.aps.org/doi/10.1103/PhysRevLett.121.053601}
}

@article{PRL2018QN,
  title = {{Quantum Synchronization and Entanglement Generation}},
  author = {Roulet, Alexandre and Bruder, Christoph},
  journal = {Phys. Rev. Lett.},
  volume = {121},
  issue = {6},
  pages = {063601},
  numpages = {5},
  year = {2018},
  month = {Aug},
  publisher = {American Physical Society},
  doi = {10.1103/PhysRevLett.121.063601},
  url = {https://link.aps.org/doi/10.1103/PhysRevLett.121.063601}
}

@article{Solanki2023PRA,
  title = {Symmetries and synchronization blockade},
  author = {Solanki, Parvinder and Mehdi, Faraz Mohd and Hajdu\v{s}ek, Michal and Vinjanampathy, Sai},
  journal = {Phys. Rev. A},
  volume = {108},
  issue = {2},
  pages = {022216},
  numpages = {9},
  year = {2023},
  month = {Aug},
  publisher = {American Physical Society},
  doi = {10.1103/PhysRevA.108.022216},
  url = {https://link.aps.org/doi/10.1103/PhysRevA.108.022216}
}

@article{PRA2024Tobia,
  title = {Quantum synchronization through the interference blockade},
  author = {Kehrer, Tobias and Nadolny, Tobias and Bruder, Christoph},
  journal = {Phys. Rev. A},
  volume = {110},
  issue = {4},
  pages = {042203},
  numpages = {10},
  year = {2024},
  month = {Oct},
  publisher = {American Physical Society},
  doi = {10.1103/PhysRevA.110.042203},
  url = {https://link.aps.org/doi/10.1103/PhysRevA.110.042203}
}

@article{PRA2025QSandQFI,
  title = {Quantum synchronization and dissipative quantum sensing},
  author = {Vaidya, Gaurav M. and J\"ager, Simon B. and Shankar, Athreya},
  journal = {Phys. Rev. A},
  volume = {111},
  issue = {1},
  pages = {012410},
  numpages = {12},
  year = {2025},
  month = {Jan},
  publisher = {American Physical Society},
  doi = {10.1103/PhysRevA.111.012410},
  url = {https://link.aps.org/doi/10.1103/PhysRevA.111.012410}
}

@article{PRA2020two,
  title = {Synchronization in two-level quantum systems},
  author = {Parra-L\'opez, \'Alvaro and Bergli, Joakim},
  journal = {Phys. Rev. A},
  volume = {101},
  issue = {6},
  pages = {062104},
  numpages = {7},
  year = {2020},
  month = {Jun},
  publisher = {American Physical Society},
  doi = {10.1103/PhysRevA.101.062104},
  url = {https://link.aps.org/doi/10.1103/PhysRevA.101.062104}
}

@article{PRL2020exp,
  title = {{Observation of Quantum Phase Synchronization in Spin-1 Atoms}},
  author = {Laskar, Arif Warsi and Adhikary, Pratik and Mondal, Suprodip and Katiyar, Parag and Vinjanampathy, Sai and Ghosh, Saikat},
  journal = {Phys. Rev. Lett.},
  volume = {125},
  issue = {1},
  pages = {013601},
  numpages = {5},
  year = {2020},
  month = {Jul},
  publisher = {American Physical Society},
  doi = {10.1103/PhysRevLett.125.013601},
  url = {https://link.aps.org/doi/10.1103/PhysRevLett.125.013601}
}

@article{PRL2018VdP,
  title = {{Squeezing Enhances Quantum Synchronization}},
  author = {Sonar, Sameer and Hajdu\v{s}ek, Michal and Mukherjee, Manas and Fazio, Rosario and Vedral, Vlatko and Vinjanampathy, Sai and Kwek, Leong-Chuan},
  journal = {Phys. Rev. Lett.},
  volume = {120},
  issue = {16},
  pages = {163601},
  numpages = {5},
  year = {2018},
  month = {Apr},
  publisher = {American Physical Society},
  doi = {10.1103/PhysRevLett.120.163601},
  url = {https://link.aps.org/doi/10.1103/PhysRevLett.120.163601}
}

@article{PRB2009QED,
  title = {Quantum synchronization and entanglement of two qubits coupled to a driven dissipative resonator},
  author = {Zhirov, O. V. and Shepelyansky, D. L.},
  journal = {Phys. Rev. B},
  volume = {80},
  issue = {1},
  pages = {014519},
  numpages = {4},
  year = {2009},
  month = {Jul},
  publisher = {American Physical Society},
  doi = {10.1103/PhysRevB.80.014519},
  url = {https://link.aps.org/doi/10.1103/PhysRevB.80.014519}
}

@article{PRB2024No_Go,
  title = {Absence of correlations in dissipative interacting qubits: A no-go theorem},
  author = {Wang, Zeqing and Qi, Ran and Lu, Yao and Wu, Zhigang and Jie, Jianwen},
  journal = {Phys. Rev. B},
  volume = {110},
  issue = {15},
  pages = {155129},
  numpages = {13},
  year = {2024},
  month = {Oct},
  publisher = {American Physical Society},
  doi = {10.1103/PhysRevB.110.155129},
  url = {https://link.aps.org/doi/10.1103/PhysRevB.110.155129}
}

@article{Zhang2023PRR,
  title = {Quantum synchronization of a single trapped-ion qubit},
  author = {Zhang, Liyun and Wang, Zhao and Wang, Yucheng and Zhang, Junhua and Wu, Zhigang and Jie, Jianwen and Lu, Yao},
  journal = {Phys. Rev. Res.},
  volume = {5},
  issue = {3},
  pages = {033209},
  numpages = {16},
  year = {2023},
  month = {Sep},
  publisher = {American Physical Society},
  doi = {10.1103/PhysRevResearch.5.033209},
  url = {https://link.aps.org/doi/10.1103/PhysRevResearch.5.033209}
}

@article{PRR2020SPin1,
  title = {{Quantum synchronization on the IBM Q system}},
  author = {Koppenh\"ofer, Martin and Bruder, Christoph and Roulet, Alexandre},
  journal = {Phys. Rev. Research},
  volume = {2},
  issue = {2},
  pages = {023026},
  numpages = {8},
  year = {2020},
  month = {Apr},
  publisher = {American Physical Society},
  doi = {10.1103/PhysRevResearch.2.023026},
  url = {https://link.aps.org/doi/10.1103/PhysRevResearch.2.023026}
}

@article{PRA2022nuclear,
  title = {Observation of quantum phase synchronization in a nuclear-spin system},
  author = {Krithika, V. R. and Solanki, Parvinder and Vinjanampathy, Sai and Mahesh, T. S.},
  journal = {Phys. Rev. A},
  volume = {105},
  issue = {6},
  pages = {062206},
  numpages = {9},
  year = {2022},
  month = {Jun},
  publisher = {American Physical Society},
  doi = {10.1103/PhysRevA.105.062206},
  url = {https://link.aps.org/doi/10.1103/PhysRevA.105.062206}
}

@article{PRE2024Sudler,
  title = {Driven generalized quantum Rayleigh--van der Pol oscillators: Phase localization and spectral response},
  author = {Sudler, A. J. and Talukdar, J. and Blume, D.},
  journal = {Phys. Rev. E},
  volume = {109},
  issue = {5},
  pages = {054207},
  numpages = {17},
  year = {2024},
  month = {May},
  publisher = {American Physical Society},
  doi = {10.1103/PhysRevE.109.054207},
  url = {https://link.aps.org/doi/10.1103/PhysRevE.109.054207}
}

@article{PRL2013OM1,
  title = {Photonic Cavity Synchronization of Nanomechanical Oscillators},
  author = {Bagheri, Mahmood and Poot, Menno and Fan, Linran and Marquardt, Florian and Tang, Hong X.},
  journal = {Phys. Rev. Lett.},
  volume = {111},
  issue = {21},
  pages = {213902},
  numpages = {5},
  year = {2013},
  month = {Nov},
  publisher = {American Physical Society},
  doi = {10.1103/PhysRevLett.111.213902},
  url = {https://link.aps.org/doi/10.1103/PhysRevLett.111.213902}
}

@article{PRA2018coldatom,
  title = {Synchronization of a self-sustained cold-atom oscillator},
  author = {Heimonen, H. and Kwek, L. C. and Kaiser, R. and Labeyrie, G.},
  journal = {Phys. Rev. A},
  volume = {97},
  issue = {4},
  pages = {043406},
  numpages = {5},
  year = {2018},
  month = {Apr},
  publisher = {American Physical Society},
  doi = {10.1103/PhysRevA.97.043406},
  url = {https://link.aps.org/doi/10.1103/PhysRevA.97.043406}
}

@article{PRA2015QED,
  title = {Mutual information as an order parameter for quantum synchronization},
  author = {Ameri, V. and Eghbali-Arani, M. and Mari, A. and Farace, A. and Kheirandish, F. and Giovannetti, V. and Fazio, R.},
  journal = {Phys. Rev. A},
  volume = {91},
  issue = {1},
  pages = {012301},
  numpages = {6},
  year = {2015},
  month = {Jan},
  publisher = {American Physical Society},
  doi = {10.1103/PhysRevA.91.012301},
  url = {https://link.aps.org/doi/10.1103/PhysRevA.91.012301}
}

@article{Weiss_2016,
	doi = {10.1088/1367-2630/18/1/013043},
	url = {https://doi.org/10.1088/1367-2630/18/1/013043},
	year = 2016,
	month = {jan},
	publisher = {{IOP} Publishing},
	volume = {18},
	number = {1},
	pages = {013043},
	author = {Talitha Weiss and Andreas Kronwald and Florian Marquardt},
	title = {Noise-induced transitions in optomechanical synchronization},
	journal = {New J. Phys.}
}

@article{PRE2012OM,
  title = {Synchronization of many nanomechanical resonators coupled via a common cavity field},
  author = {Holmes, C. A. and Meaney, C. P. and Milburn, G. J.},
  journal = {Phys. Rev. E},
  volume = {85},
  issue = {6},
  pages = {066203},
  numpages = {15},
  year = {2012},
  month = {Jun},
  publisher = {American Physical Society},
  doi = {10.1103/PhysRevE.85.066203},
  url = {https://link.aps.org/doi/10.1103/PhysRevE.85.066203}
}

@article{PRA2014OM,
  title = {Quantum manifestation of a synchronization transition in optomechanical systems},
  author = {Ying, Lei and Lai, Ying-Cheng and Grebogi, Celso},
  journal = {Phys. Rev. A},
  volume = {90},
  issue = {5},
  pages = {053810},
  numpages = {6},
  year = {2014},
  month = {Nov},
  publisher = {American Physical Society},
  doi = {10.1103/PhysRevA.90.053810},
  url = {https://link.aps.org/doi/10.1103/PhysRevA.90.053810}
}

@article{PRL2012NOexp,
  title = {{Synchronization of Micromechanical Oscillators Using Light}},
  author = {Zhang, Mian and Wiederhecker, Gustavo S. and Manipatruni, Sasikanth and Barnard, Arthur and McEuen, Paul and Lipson, Michal},
  journal = {Phys. Rev. Lett.},
  volume = {109},
  issue = {23},
  pages = {233906},
  numpages = {5},
  year = {2012},
  month = {Dec},
  publisher = {American Physical Society},
  doi = {10.1103/PhysRevLett.109.233906},
  url = {https://link.aps.org/doi/10.1103/PhysRevLett.109.233906}
}

@incollection{galve2017quantum,
  title={Quantum correlations and synchronization measures},
  author={Galve, Fernando and Luca Giorgi, Gian and Zambrini, Roberta},
  booktitle={Lectures on general quantum correlations and their applications},
  pages={393--420},
  year={2017},
  publisher={Springer}
}

@article{PRA2013Spin,
  title = {Spontaneous synchronization and quantum correlation dynamics of open spin systems},
  author = {Giorgi, G. L. and Plastina, F. and Francica, G. and Zambrini, R.},
  journal = {Phys. Rev. A},
  volume = {88},
  issue = {4},
  pages = {042115},
  numpages = {9},
  year = {2013},
  month = {Oct},
  publisher = {American Physical Society},
  doi = {10.1103/PhysRevA.88.042115},
  url = {https://link.aps.org/doi/10.1103/PhysRevA.88.042115}
}

@article{he_entanglement_2024,
author = {Si-Wen He and Zhi Jiao Deng and Yi Xie and Yan-Yi Wang and Ping-Xing Chen},
journal = {Opt. Express},
number = {8},
pages = {13998--14009},
publisher = {Optica Publishing Group},
title = {Entanglement signatures for quantum synchronization with single-ion phonon laser},
volume = {32},
month = {Apr},
year = {2024},
url = {https://opg.optica.org/oe/abstract.cfm?URI=oe-32-8-13998},
doi = {10.1364/OE.515903},
}

@article{shen_fisher_2023,
author = {Shen, Yuan and Soh, Hong Yi and Kwek, Leong-Chuan and Fan, Weijun},
journal = {Entropy},
number = {8},
pages = {1116},
title = {Fisher Information as General Metrics of Quantum Synchronization},
volume = {25},
year = {2023},
url = {https://www.mdpi.com/1099-4300/25/8/1116},
doi = {10.3390/e25081116},
}

@article{wachtler2023topological,
  title = {Topological synchronization of quantum van der Pol oscillators},
  author = {W\"achtler, Christopher W. and Platero, Gloria},
  journal = {Phys. Rev. Res.},
  volume = {5},
  issue = {2},
  pages = {023021},
  numpages = {18},
  year = {2023},
  month = {Apr},
  publisher = {American Physical Society},
  doi = {10.1103/PhysRevResearch.5.023021},
  url = {https://link.aps.org/doi/10.1103/PhysRevResearch.5.023021}
}

@article{wachtler2024topological,
  title = {Topological Quantum Synchronization of Fractionalized Spins},
  author = {W\"achtler, Christopher W. and Moore, Joel E.},
  journal = {Phys. Rev. Lett.},
  volume = {132},
  issue = {19},
  pages = {196601},
  numpages = {7},
  year = {2024},
  month = {May},
  publisher = {American Physical Society},
  doi = {10.1103/PhysRevLett.132.196601},
  url = {https://link.aps.org/doi/10.1103/PhysRevLett.132.196601}
}

@article{mr1f-v8cv,
  title = {Quantum synchronization in one-dimensional topological systems},
  author = {Liu, Tong and Garc\'{\i}a-\'Alvarez, Laura and Tancredi, Giovanna},
  journal = {Phys. Rev. Res.},
  volume = {7},
  issue = {2},
  pages = {L022064},
  numpages = {7},
  year = {2025},
  month = {Jun},
  publisher = {American Physical Society},
  doi = {10.1103/mr1f-v8cv},
  url = {https://link.aps.org/doi/10.1103/mr1f-v8cv}
}

@article{Yang2025,
  author    = {Shu Yang and Zeqing Wang and Libin Fu and Jianwen Jie},
  title     = {Emergent continuous time crystal in dissipative quantum spin system without driving},
  journal   = {Communications Physics},
  year      = {2025},
  volume    = {8},
  number    = {1},
  pages     = {114},
  doi       = {10.1038/s42005-025-02040-1},
  url       = {https://doi.org/10.1038/s42005-025-02040-1},
  issn      = {2399-3650}
}

@article{PRB2023Wang,
  title = {Accelerating relaxation dynamics in open quantum systems with Liouvillian skin effect},
  author = {Wang, Zeqing and Lu, Yao and Peng, Yi and Qi, Ran and Wang, Yucheng and Jie, Jianwen},
  journal = {Phys. Rev. B},
  volume = {108},
  issue = {5},
  pages = {054313},
  numpages = {8},
  year = {2023},
  month = {Aug},
  publisher = {American Physical Society},
  doi = {10.1103/PhysRevB.108.054313},
  url = {https://link.aps.org/doi/10.1103/PhysRevB.108.054313}
}

@article{PRL2014VdP,
  title = {Quantum Synchronization of a Driven Self-Sustained Oscillator},
  author = {Walter, Stefan and Nunnenkamp, Andreas and Bruder, Christoph},
  journal = {Phys. Rev. Lett.},
  volume = {112},
  issue = {9},
  pages = {094102},
  numpages = {5},
  year = {2014},
  month = {Mar},
  publisher = {American Physical Society},
  doi = {10.1103/PhysRevLett.112.094102},
  url = {https://link.aps.org/doi/10.1103/PhysRevLett.112.094102}
}

@article{PRL2013VdP,
  title = {Quantum Synchronization of Quantum van der Pol Oscillators with Trapped Ions},
  author = {Lee, Tony E. and Sadeghpour, H. R.},
  journal = {Phys. Rev. Lett.},
  volume = {111},
  issue = {23},
  pages = {234101},
  numpages = {5},
  year = {2013},
  month = {Dec},
  publisher = {American Physical Society},
  doi = {10.1103/PhysRevLett.111.234101},
  url = {https://link.aps.org/doi/10.1103/PhysRevLett.111.234101}
}

@article{PRL2019VdP,
  title = {Critical Response of a Quantum van der Pol Oscillator},
  author = {Dutta, Shovan and Cooper, Nigel R.},
  journal = {Phys. Rev. Lett.},
  volume = {123},
  issue = {25},
  pages = {250401},
  numpages = {6},
  year = {2019},
  month = {Dec},
  publisher = {American Physical Society},
  doi = {10.1103/PhysRevLett.123.250401},
  url = {https://link.aps.org/doi/10.1103/PhysRevLett.123.250401}
}

@article{Tan2022halfintegervs,
  doi = {10.22331/q-2022-12-29-885},
  url = {https://doi.org/10.22331/q-2022-12-29-885},
  title = {Half-integer vs. integer effects in quantum synchronization of spin systems},
  author = {Tan, Ryan and Bruder, Christoph and Koppenh{\"{o}}fer, Martin},
  journal = {{Quantum}},
  issn = {2521-327X},
  publisher = {{Verein zur F{\"{o}}rderung des Open Access Publizierens in den Quantenwissenschaften}},
  volume = {6},
  pages = {885},
  month = dec,
  year = {2022}
}

@article{PRR2020hybrid,
  title = {Generalized measure of quantum synchronization},
  author = {Jaseem, Noufal and Hajdu\v{s}ek, Michal and Solanki, Parvinder and Kwek, Leong-Chuan and Fazio, Rosario and Vinjanampathy, Sai},
  journal = {Phys. Rev. Res.},
  volume = {2},
  issue = {4},
  pages = {043287},
  numpages = {8},
  year = {2020},
  month = {Nov},
  publisher = {American Physical Society},
  doi = {10.1103/PhysRevResearch.2.043287},
  url = {https://link.aps.org/doi/10.1103/PhysRevResearch.2.043287}
}

@article{NC2025lai,
  title     = {Nonreciprocal quantum synchronization},
  author    = {Deng-Gao Lai and Adam Miranowicz and Franco Nori},
  journal   = {Nature Communications},
  year      = {2025},
  volume    = {16},
  number    = {1},
  pages     = {8491},
  doi       = {10.1038/s41467-025-63408-z}
}

@article{PRL2025Model,
  title = {Quantum Origin of Limit Cycles, Fixed Points, and Critical Slowing Down},
  author = {Dutta, Shovan and Zhang, Shu and Haque, Masudul},
  journal = {Phys. Rev. Lett.},
  volume = {134},
  issue = {5},
  pages = {050407},
  numpages = {7},
  year = {2025},
  month = {Feb},
  publisher = {American Physical Society},
  doi = {10.1103/PhysRevLett.134.050407},
  url = {https://link.aps.org/doi/10.1103/PhysRevLett.134.050407}
}

@article{PRL2018BTC,
  title = {Boundary Time Crystals},
  author = {Iemini, F. and Russomanno, A. and Keeling, J. and Schir\`o, M. and Dalmonte, M. and Fazio, R.},
  journal = {Phys. Rev. Lett.},
  volume = {121},
  issue = {3},
  pages = {035301},
  numpages = {6},
  year = {2018},
  month = {Jul},
  publisher = {American Physical Society},
  doi = {10.1103/PhysRevLett.121.035301},
  url = {https://link.aps.org/doi/10.1103/PhysRevLett.121.035301}
}

@article{RMP2023Norman,
  title = {Colloquium: Quantum and classical discrete time crystals},
  author = {Zaletel, Michael P. and Lukin, Mikhail and Monroe, Christopher and Nayak, Chetan and Wilczek, Frank and Yao, Norman Y.},
  journal = {Rev. Mod. Phys.},
  volume = {95},
  issue = {3},
  pages = {031001},
  numpages = {34},
  year = {2023},
  month = {Jul},
  publisher = {American Physical Society},
  doi = {10.1103/RevModPhys.95.031001},
  url = {https://link.aps.org/doi/10.1103/RevModPhys.95.031001}
}

@article{PRL2012CTC,
  title = {Classical Time Crystals},
  author = {Shapere, Alfred and Wilczek, Frank},
  journal = {Phys. Rev. Lett.},
  volume = {109},
  issue = {16},
  pages = {160402},
  numpages = {4},
  year = {2012},
  month = {Oct},
  publisher = {American Physical Society},
  doi = {10.1103/PhysRevLett.109.160402},
  url = {https://link.aps.org/doi/10.1103/PhysRevLett.109.160402}
}

@article{PRL2012QTC,
  title = {Quantum Time Crystals},
  author = {Wilczek, Frank},
  journal = {Phys. Rev. Lett.},
  volume = {109},
  issue = {16},
  pages = {160401},
  numpages = {5},
  year = {2012},
  month = {Oct},
  publisher = {American Physical Society},
  doi = {10.1103/PhysRevLett.109.160401},
  url = {https://link.aps.org/doi/10.1103/PhysRevLett.109.160401}
}

@article{PRB2021dbtc,
  title = {Boundary time crystals in collective $d$-level systems},
  author = {Prazeres, Luis Fernando dos and Souza, Leonardo da Silva and Iemini, Fernando},
  journal = {Phys. Rev. B},
  volume = {103},
  issue = {18},
  pages = {184308},
  numpages = {16},
  year = {2021},
  month = {May},
  publisher = {American Physical Society},
  doi = {10.1103/PhysRevB.103.184308},
  url = {https://link.aps.org/doi/10.1103/PhysRevB.103.184308}
}

@article{Montenegro2023,
  title     = {Quantum metrology with boundary time crystals},
  author    = {Victor Montenegro and Marco G. Genoni and Abolfazl Bayat and Matteo G. A. Paris},
  journal   = {Communications Physics},
  year      = {2023},
  volume    = {6},
  number    = {1},
  pages     = {304},
  doi       = {10.1038/s42005-023-01423-6},
  url       = {https://doi.org/10.1038/s42005-023-01423-6},
  issn      = {2399-3650}
}

@article{QVDP_2025_Lin,
author = {Yi Li  and Zihan Xie  and Xiaodong Yang  and Yue Li  and Xingyu Zhao  and Xu Cheng  and Xinhua Peng  and Jun Li  and Eric Lutz  and Yiheng Lin  and Jiangfeng Du },
title = {Experimental realization and synchronization of a quantum van der Pol oscillator},
journal = {Science Advances},
volume = {11},
number = {41},
pages = {eady5649},
year = {2025},
doi = {10.1126/sciadv.ady5649},
URL = {https://www.science.org/doi/abs/10.1126/sciadv.ady5649},
eprint = {https://www.science.org/doi/pdf/10.1126/sciadv.ady5649}}

@article{46my-41ym,
  title = {Universal interaction-based manipulation of quantum synchronization in spin oscillator networks},
  author = {Dai, Shuo and Wang, Zeqing and Wan, Liang-Liang and Li, Weidong and Smerzi, Augusto and Qi, Ran and Jie, Jianwen},
  journal = {Phys. Rev. B},
  volume = {113},
  pages = {054306},
  year = {2026},
  doi = {10.1103/46my-41ym},
  url = {https://doi.org/10.1103/46my-41ym}
}

@book{Kuznetsov2004,
  author = {Kuznetsov, Yuri A.},
  title = {Elements of Applied Bifurcation Theory},
  edition = {3},
  year = {2004},
  publisher = {Springer},
  address = {New York},
  series = {Applied Mathematical Sciences},
  volume = {112},
  doi = {10.1007/978-1-4757-3978-7},
  url = {https://doi.org/10.1007/978-1-4757-3978-7}
}

@article{Wang2026,
  title = {Non-Resonant Boundary Time Crystals from Quantum Synchronization Breakdown},
  author = {Wang, Jun and Yang, Shu and Wang, Zeqing and Qi, Ran and Hu, Haiping and Li, Weidong and Jie, Jianwen},
  journal = {arXiv:2603.14311},
  year = {2026},
  url = {https://arxiv.org/abs/2603.14311}
}

@article{Jie2026SingleSpin,
  author  = {Jie, Jianwen},
  title   = {A Single Spin Switches the Steady-State Phase of an Open Quantum System},
  journal = {arXiv preprint arXiv:2608.28390},
  year    = {2026},
  url     = {https://arxiv.org/abs/2608.28390}
}

@article{Arinushkin2021NonlinearDamping,
  title   = {Nonlinear damping effects in a simplified power grid model based on coupled {Kuramoto}-like oscillators with inertia},
  author  = {Arinushkin, P. A. and Vadivasova, T. E.},
  journal = {Chaos, Solitons \& Fractals},
  volume  = {152},
  pages   = {111343},
  year    = {2021},
  doi     = {10.1016/j.chaos.2021.111343}
}

@article{Liu2026SolitonSynchronization,
  title   = {Synchronization and bifurcation dynamics of dissipative breathing solitons in pure-quartic fiber lasers via external modulation},
  author  = {Liu, Chunting and Liu, Hengyu and Shen, Hanyang and Cai, Yu and Qiang, Zong and Zhang, Zuxing},
  journal = {Chaos, Solitons \& Fractals},
  volume  = {208},
  pages   = {118164},
  year    = {2026},
  doi     = {10.1016/j.chaos.2026.118164}
}

@article{Duan2026PumpControlled,
  title   = {Pump-controlled bifurcation cascade in a dissipative soliton fiber laser: From fixed points to attractor competition},
  author  = {Duan, Xinxu and Liu, Yuantong and Tang, Xiaoyun and Jiang, Hongbo and Jin, Lei},
  journal = {Chaos, Solitons \& Fractals},
  volume  = {208},
  pages   = {118311},
  year    = {2026},
  doi     = {10.1016/j.chaos.2026.118311}
}
\end{document}